\documentclass[letterpaper]{aa}
\usepackage{natbib}
\bibpunct{(}{)}{;}{a}{}{,}
\usepackage{graphicx}

\begin{document}
\title{A kinetic control of the heliospheric interface hydrodynamics of
charge-exchanging fluids}
\author{H.J. Fahr \inst{1} \and M. Bzowski \inst{2}}
\offprints{H. J. Fahr, \email{hfahr@astro.uni-bonn.de}}
\institute{Insitute of Astrophysics and Extraterrestrial Research,
Bonn University, Auf dem H{\"u}gel 71, Bonn, Germany
\and Space Research Centre PAS, Bartycka 18A, Warsaw, Poland}
\date{Received 26 October 2003 / Accepted 27 October 2003}

\newcommand{\asr}{Adv. Space Res.}
\newcommand{\ag}{Ann. Geophys.}
\newcommand{\pss}{Planet. Space Sci.}
\abstract{

It is well known that the Solar System is presently moving through a
partially ionized local interstellar medium. This gives rise to a
counter-flow situation requiring a consistent description of behaviour
of the two fluids -- ions and neutral atoms -- which are dynamically
coupled by mutual charge exchange processes. Solutions to this problem
have been offered in the literature, all relying on the assumption that
the proton fluid, even under evidently nonequilibrium conditions, can be
expected to stay in a highly-relaxated distribution function given by
mono-Maxwellians shifted by the local proton bulk velocity. Here we
check the validity of this assumption, calculating on the basis of a
Boltzmann-kinetic approach the actually occurring deviations. As we
show, especially for low degrees of ionization, $\xi \le 0.3$, both the
H-atoms and protons involved do generate in the heliospheric interface
clearly pronounced deviations from shifted Maxwellians with
asymmetrically shaped distribution functions giving rise to
non-convective transport processes and heat conduction flows. Also in
the inner heliosheath region and in the heliotail deviations of the
proton distribution from the hydrodynamic one must be expected. This
sheds new light on the correctness of current calculations of H-atom
distribution functions prevailing in the inner heliosphere and also of
the Lyman-$\alpha $ absorption features in stellar spectra due to the
presence of the hydrogen wall atoms. Deviations from LTE-functions would
be even more pronounced in magnetic interfaces, which via CGL-effects
cause temperature anisotropies to arise. }

\titlerunning{Kinetic charge exchange coupling in HD heliospheric interface} 

\authorrunning{H.J. Fahr \& M.Bzowski} 
\maketitle

\section{Introduction}

The problem of the heliospheric interface, where H-atoms and protons are
effectively coupled by charge exchange interactions, has often been
faced in the literature. In general, it was recognized very early that
the passage of neutral interstellar atoms (O, H) through the plasma
interface ahead of the solar wind termination shock needs a kinetic
treatment, since the relevant charge exchange mean free paths between
H-atoms and protons are comparable to or even larger than the typical
structure scales of the interface plasma flow (i.e., Knudsen numbers are
equal to or smaller than 1; see Ripken \& Fahr
1983\nocite{ripken_fahr:83a}; Fahr \& Ripken
1984\nocite{fahr_ripken:84}; Fahr 1991\nocite{fahr:91}; Osterbart \&
Fahr 1992\nocite{osterbart_fahr:92}; Baranov \& Malama 1993
\nocite{baranov_malama:93}; McNutt et al., 1998\nocite{mcnutt_etal:98},
Bzowski et al. 2000\nocite{bzowski_etal:00}; Izmodenov et al. 2001)
\nocite{izmodenov_etal:01a}. Nevertheless, due to mathematical
complications associated with such kinetic treatments of the problem,
many heliospheric models have appeared in the literature which use
hydrodynamic treatments of the two fluids, H-atoms and protons, coupled
by charge exchange reactions (for recent reviews, see Zank
1999;\nocite{zank:99} Fahr 2003\nocite{fahr:03c}b).

The hydrodynamics applied in all these approaches is restricted to the
description of the space-time behavior of the lowest hydrodynamic
moments like density, bulk velocity and pressure and the local
distribution function is taken to be solely a function of these moments
in the form of shifted Maxwellians. As suggested in Fig.\ref{figDemo},
the permanent supply of newly charge-exchanged particles into the
local distribution functions will maintain the resulting distribution
(shown by dashed lines) far from a three-moment HD distribution.

This has been clearly recognized by Baranov and Malama (1993,
1995)\nocite{baranov_malama:93}\nocite{baranov_malama:95}, who for this
reasons perfected a kinetic treatment for the neutral H-atoms, even in
view of the mathematical complications. The semi-kinetic approach
offered by them, though treating the H-atom kinetically, is still based
on the assumption that the protons can be described as a hydrodynamic
fluid. But this is not true under special conditions, as we will
demonstrate here. The protons generate deviations from a 3-moment
HD-distribution. Therefore to represent the two fluids interacting by
charge exchange reactions one would need a kinetic description both for
the H-atoms and the protons. This highly complicated model will not be
offered in this paper, but we present calculations which clearly make
visible the resulting deviations of H-atom and proton distributions from
3-moment HD distributions. One other point, which was already emphasized
by Fahr (2003a,b),\nocite{fahr:03a}\nocite{fahr:03b} concerns the fact
that hydrodynamic two-fluid descriptions of the interface flows use
charge-exchange coupling terms which are only justified in cases of
supersonic bulk velocity differences. Under realistic interface
conditions the resulting bulk velocity differences, however, appear to
have subsonic magnitudes and thus require revised formulations of the
charge exchange coupling terms. In the following we thus subject
hydrodynamic approaches to a kinetic control.

\section{Theoretical approach}
\begin{figure}
\resizebox{\hsize}{!}{\includegraphics{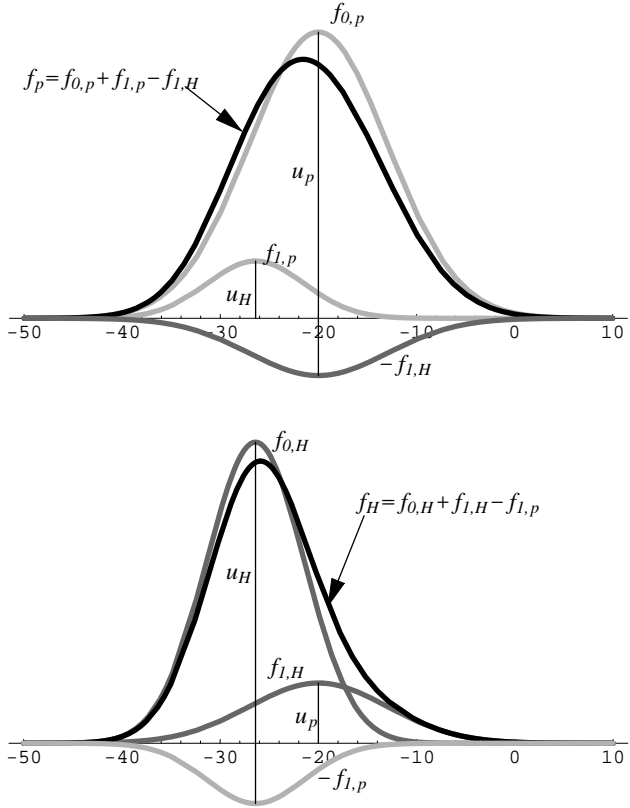}}
\caption{Schematic view of the ``old'' proton and H-atom distribution
functions ($f_{0,p}$ and $f_{0,H}$, light gray and dark gray, respectively),
with different bulk velocities $u_p$, $u_H$ and temperatures, interacting
with each other by
charge exchange. The charge exchange gives rise to secondary populations
of protons and H-atoms that inherit the kinematic properties of their
respective source populations of unperturbed H-atoms and protons. As a result,
non-Maxwellian distribution functions $f_p$, $f_H$ of both protons and
H-atoms appear, shown in  black lines.}
\label{figDemo}
\end{figure}
\begin{figure}
\resizebox{\hsize}{!}{\includegraphics{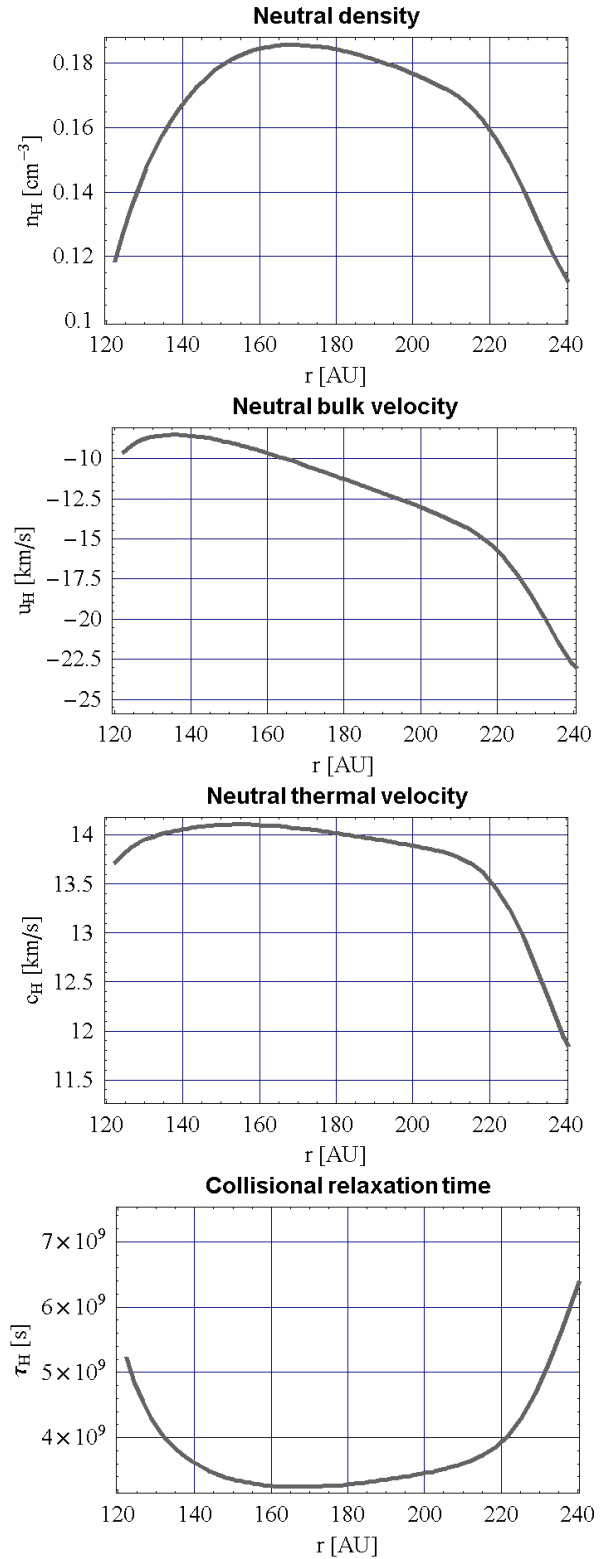}}
\caption{Macroscopic parameters of H-atoms between the bow shock and the
heliopause, calculated with the use of the multi-fluid hydrodynamical treatment
developed by Fahr et al. (2000). The bow shock is
at $\sim 250$~AU, the heliopause at $\sim 125$~AU.}
\label{figHParam}
\end{figure}
\begin{figure}
\resizebox{\hsize}{!}{\includegraphics{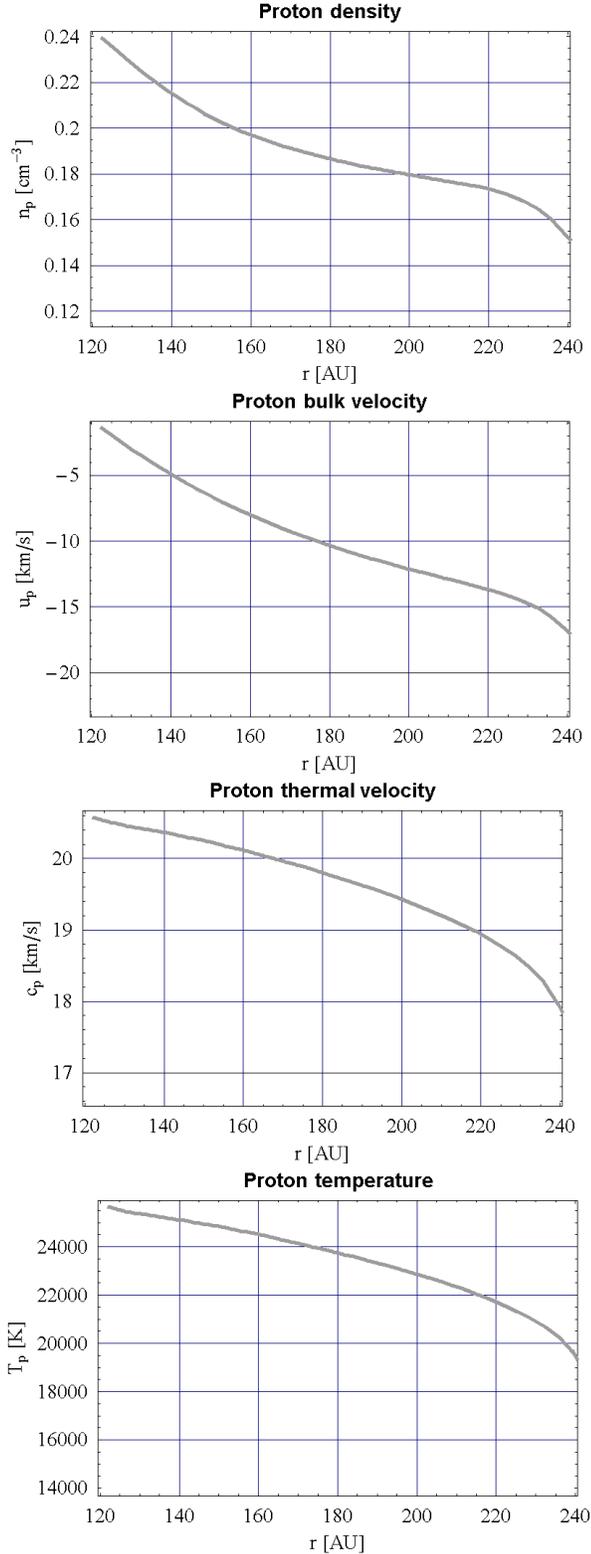}}
\caption{Macroscopic parameters of protons between the bow shock and the
heliopause, calculated with the use of the multi-fluid hydrodynamical treatment
developed by Fahr et al. (2000). }
\label{figPParam}
\end{figure}

We assume that protons and hydrogen atoms in the interface are coupled
solely by charge exchange processes. For simplicity, we limit ourselves
to calculations on the stagnation line only. Denoting the hydrogen
distribution function by $f_{H}$ and the proton distribution function by
$f_{p}$ we furthermore assume that the two distribution functions can be
represented by Maxwellian cores $f_{0,H}$, $ f_{0,p}$ from hydrodynamic
calculations and by non-relaxed corrections $f_{1,H}$, $f_{1,p}$ due to
mutual implantations of new charge exchanged particles, as shown in the
following formulae and qualitatively in Fig.\ref{figDemo}:
\begin{equation}
f_{H}\left( \vec{v},s\right) =f_{0,H}\left( \vec{v},s\right) +f_{1,H}\left(
\vec{v},s\right)  \label{eqn7a}
\end{equation}
\begin{equation}
f_{p}\left( \vec{v},s\right) =f_{0,p}\left( \vec{v},s\right) +f_{1,p}\left(
\vec{v},s\right)  \label{eqn7b}
\end{equation}
where $s$ measures the spatial position on the symmetry axis. At this
symmetry axis, and for stationary conditions, we have the following system
of Boltzmann-kinetic equations in phase space, which couple the protons and
hydrogen atoms via mutual losses and gains (see, e.g., Fahr
1996\nocite{fahr:96}):
\begin{equation}
v\cos \theta ~\frac{df_{H}}{ds} =\frac{\delta ^{+}f_{H}}{\delta t}-\frac{
\delta ^{-}f_{H}}{\delta t}+\frac{f_{0,H}-f_{H}}{\tau _{H}}  \label{eqn1}
\end{equation}
\begin{equation}
v\cos \theta ~\frac{df_{p}}{ds} =\frac{\delta ^{+}f_{p}}{\delta t}-\frac{
\delta ^{-}f_{p}}{\delta t}+\frac{f_{0,p}-f_{p}}{\tau _{p}}  \label{eqn2}
\end{equation}
In these equations $s$ is a line element along the axis in the interface
and $\tau _{H}$, $\tau _{p}$ are relaxation times, separate for protons
and hydrogen atoms and dependent on $s$ and the particle velocities. The
terms with $\delta^+$ and $\delta^-$ are, correspondingly, the
production and loss terms, as shown in formulae \ref{eqn5} and
\ref{eqn6}, and the remaining terms in Eqs \ref{eqn1} and \ref{eqn2} are
the relaxation terms operating to restitute the hydrodynamic solutions
for appropriate distribution functions. The actually resulting
distribution function $f$ which, had the local charge exchange
influences been stopped, would relax towards the hydrodynamic core
distribution $f_0$ within one relaxation period. But as long as charge
exchange is operating, the resulting distribution is permanently kept
off this function $f_0$. In other words, we assume that two-fluid
hydrodynamics would deliver correct results in the interface had the
H-atom and proton relaxation processes operated fast enough, i.e., if
$\tau_H,\, \tau_p << \tau_{ex}$, $\tau_{ex}$ being the charge exchange
period between the H-atoms and protons. Only because $\tau_H,\, \tau_p
\simeq \tau_{ex}$ deviations from the hydrodynamic core functions are
pronounced. The above system of differential equations is valid for an
arbitrary velocity vector $\vec{v}\left( v,\cos \theta \right)$ but for
a proton velocity vector identical to the H atom velocity vector.
$\theta$ is the angle between the direction of atom's motion and the
inflow axis $z$ (and consequently $v \cos \theta = v_z$). No matter what
the forms of the distribution functions are, a phase-space hydrogen gain is
then a phase-space proton loss and vice versa. Hence equations
\ref{eqn1} and \ref{eqn2} are modified to
\begin{eqnarray}
v\cos \theta ~\frac{df_{H}}{ds} &=&\frac{\delta ^{-}f_{p}}{\delta t}-\frac{
\delta ^{-}f_{H}}{\delta t}-\frac{f_{1,H}}{\tau _{H}}  \label{eqn3}
\end{eqnarray}
\begin{eqnarray}
v\cos \theta ~\frac{df_{p}}{ds} &=&\frac{\delta ^{-}f_{H}}{\delta t}-\frac{
\delta ^{-}f_{p}}{\delta t}-\frac{f_{1,p}}{\tau _{p}}  \label{eqn4}
\end{eqnarray}

Specifically, the two remaining coupling terms are of the following forms:
\begin{eqnarray}
\label{eqn5}
\frac{\delta ^{-}f_{H}}{\delta t} &=& f_{H}\left( v,s,\theta _{H}\right) \times  \\
                             &\times& \int v_{rel}\left( v,v_{p},\theta _{p}\right) ~f_{p}\left( s,v_{p},\theta _{p}\right) ~\sigma \left( v_{rel}\right) ~d^{3}v_{p} \nonumber
\end{eqnarray}
\begin{eqnarray}
\label{eqn6}
\frac{\delta^{-}f_{p}}{\delta t} &=&f_{p}\left( v,s,\theta _{p}\right) \times \\
                          & \times &\int v_{rel}\left( v,v_{H},\theta _{H}\right) ~f_{H}\left( s,v_{H},\theta _{H}\right) ~\sigma \left( v_{rel}\right) ~d^{3}v_{H} \nonumber
\end{eqnarray}
Here $v_{rel}\left(v, v_p, \theta_p \right)$ denotes the relative
velocity between an H-atom of velocity $v$ and a proton of velocity
$v_p$ inclined at an angle $\theta_p$ to the proton bulk speed
$\vec{u}_p$. $\sigma\left(v_{rel} \right)$ is the H-atom -- proton
charge exchange cross section for particles with a relative velocity
$v_{rel}$.

The core functions $f_{0,H/p}$ in the integrands will be taken from a
hydrodynamic multifluid simulation code (Fahr et al.
2000\nocite{fahr_etal:00}) \nocite{fahr_etal:00} describing the
interface system. They are functions of $s$ and defined as follows
\begin{eqnarray}
\label{eqn7}
f_{0,H}\left( v,s,\theta _{H}\right) &=&\left( \frac{1}{\pi ~c_{H}}\right)^{\frac{3}{2}}~n_{H} \times  \\
                              & \times & \exp \left[ -\frac{v^{2}+u_{H}^{2}-2u_{H}~v~\cos \theta
_{H}}{c_{H}^{2}~}\right]  \nonumber
\end{eqnarray}
\begin{eqnarray}
\label{eqn8}
f_{0,p}\left( v,s,\theta _{p}\right) &=&\left( \frac{1}{\pi ~c_{p}}\right)^{\frac{3}{2}}~n_{p} \times  \\
                              & \times & \exp \left[ -\frac{v^{2}+u_{p}^{2}-2u_{p}~v~\cos \theta_{p}}{c_{p}^{2}~}\right]  \nonumber
\end{eqnarray}
where $u_{H}$, $u_{p}$ are, respectively, the local bulk velocities of
hydrogen atoms and protons, which depend on the position and thus here are
given as functions of the path length $s$ measured between the outer shock
and the heliopause; $n_{H}$, $n_{p}$ are number densities, also dependent on
$s$; $\theta _{H}$, $\theta _{p}$ are angles between the local bulk
velocities, which of course on the axis must be parallel to the inflow axis,
and the individual velocity $\vec{v}$; and $c_{H}$, $c_{p}$ are thermal
velocities of the core hydrogen atoms and protons, also dependent on $s$ via
temperatures $T_H$, $T_p$:
\begin{equation}
c_{H}^{2} =\frac{2K~T_{H}\left( s\right) }{m}=\frac{2}{m}\frac{P_{H}\left(
s\right) }{n_{H}\left( s\right) }  \label{eqn9}
\end{equation}
\begin{equation}
c_{p}^{2} =\frac{2K~T_{p}\left( s\right) }{m}=\frac{2}{m}\frac{P_{p}\left(
s\right) }{n_{p}\left( s\right) }  \label{eqn10}
\end{equation}
We assume that the proton mass $m$ is equal to the hydrogen mass. The
temperatures are also functions of $s$. In Figures \ref{figHParam} and
\ref{figPParam} we present the core hydrodynamic quantities (density,
bulk velocity, thermal velocity, and temperature) as functions of $s$,
as they result from the multi-fluid interaction code developed by Fahr
et al. (2000), \nocite{fahr_etal:00} which here was run for the
following input parameters: solar wind speed: 400~km/s, solar wind
density at 1~AU: 5~cm$^{-3}$, solar wind temperature at 1~AU:
$10^{5}$~K, interstellar gas speed: 26~km/s, interstellar proton
density: 0.1~cm$^{-3}$, interstellar hydrogen atom density:
0.1~cm$^{-3}$, interstellar gas temperature: 8000~K. The influences of
pick-up ions and of anomalous and regular components of cosmic rays were
switched off in this modelling for consistency with our
kinetic approach presented in this paper, where couplings to these
particle populations are not taken into account. 

We thus start from a two-fluid hydrodynamic interface model in which
H-atoms are represented by one single fluid only. This is different from
modellings like that presented by Zank et al. (1996),
\nocite{zank_etal:96a} where neutral hydrogen is decribed by three
different fluids representing H-atoms a) coming directly from the
interstellar medium, b) originating in the heliosheath, and c)
originating in the inner heliosphere. For our approach here, where we
want to check on kinetic deviations from hydrodynamic H-atom fluid
approximations, we consider it to be more practical to start from a
simple mono-Maxwellian representation of the H-atoms, thereby making our
concept logically more convincing.

We rewrite the coupling terms defined in equations \ref{eqn5} and
\ref{eqn6} so that the core + correction form of the distribution function
(see Eq.\ref{eqn7a} and \ref{eqn7b}) is taken into account:
\begin{eqnarray}
\label{eqn11}
\frac{\delta ^{-}f_{H}}{\delta t} &=& f_{H}\left( \vec{v},s,\theta _{H}\right) \times \\
                           & \times & \left( \int v_{rel}\left( \vec{v},\vec{v}_{p}\right) f_{0,p}\left( s,\vec{v}_{p}\right) \sigma \left( v_{rel}\right) d^{3}v_{p} + \right. \nonumber \\
                                  &+& \left. \int v_{rel}\left( \vec{v},\vec{v}_{p}\right) f_{1,p}\left( s,\vec{v}_{p}\right) \sigma \left( v_{rel}\right) d^{3}v_{p}\right) \nonumber
\end{eqnarray}
\begin{eqnarray}
\label{eqn12}
\frac{\delta ^{-}f_{p}}{\delta t} &=& f_{p}\left( \vec{v},s,\theta _{p}\right) \times \\
                           & \times &\left( \int v_{rel}\left( \vec{v},\vec{v}_{H}\right) f_{0,H}\left( s,\vec{v}_{H}\right) \sigma \left( v_{rel}\right) d^{3}v_{H}  + \right. \nonumber \\
                                & + & \left. \int v_{rel}\left( \vec{v},\vec{v}_{H}\right) f_{1,H}\left( s,\vec{v}_{H}\right) \sigma \left( v_{rel}\right) d^{3}v_{H}\right) \nonumber
\end{eqnarray}
and then we neglect the correction terms in the integrals,
since they only contribute according to their partial densities with
$n_{1,H}\ll n_{0,H}$, $n_{1,p}\ll n_{0,p}$.
We assume that the total density $n_1$ given by $f_1$ is negligible with
respect to the total density $n_0$ given by $f_0$. This does not mean that
$f_1$ itself is everywhere small with respect to $f_0$, as shown in
Fig.\ref{figDemo}, where velocity regions where $f_1 \ge f_0$ can be seen. We thus
assume that in Eqs. \ref{eqn13} and \ref{eqn14} only terms
$\left(f_{0,i} + f_{1,i} \right) \int f_{0,j}... $ must be taken into account.
Hence only the shifted
Maxwellians stay in the integrands of equations \ref{eqn11} and \ref{eqn12},
yielding:
\begin{eqnarray}
\label{eqn13}
\frac{\delta ^{-}f_{H}}{\delta t} &=& f_{H}\left( \vec{v},s,\theta _{H}\right) \times \\
                           & \times & \int v_{rel}\left( \vec{v},\vec{v}_{p}\right) ~f_{0,p}\left( s,\vec{v}_{p}\right) ~\sigma \left( v_{rel}\right) ~d^{3}v_{p} \nonumber
\end{eqnarray}
\begin{eqnarray}
\label{eqn14}
\frac{\delta ^{-}f_{p}}{\delta t} &=& f_{p}\left( \vec{v},s,\theta _{p}\right)  \times \\
                           & \times & \int v_{rel}\left( \vec{v},\vec{v}_{H}\right) ~f_{0,H}\left( s,\vec{v}_{H}\right) ~\sigma \left( v_{rel}\right) ~d^{3}v_{H} \nonumber
\end{eqnarray}

The validity of this assumption can be checked aposteriori in the
results presented in Figures \ref{figSoluNoRelax} through
\ref{figSoluRelaxRel}. It can be seen that close to the heliopause this
assumption is violated.

The dependence of the charge exchange cross section on the relative velocity
of colliding particles can be described by the well-known formula given by
Maher \& Tinsley (1977)\nocite{maher_tinsley:77}:
\begin{equation}
\sigma (v_{rel})=(A+B\log (v_{rel}))^{2},  \label{eqn14a}
\end{equation}
where $A = 1.6\, 10^{-7}$ and $B = -6.8\,10^{-9}$.
Now we further simplify, by assuming that the charge exchange cross
section can be expanded into a constant +\ linear expansion term, i.e.,
expanding around the bulk velocity $\vec{u}_p$ of the protons and bulk velocity $\vec{u}_H$
of the H-atoms, respectively:
\begin{eqnarray}
\sigma _{p}\left( \vec{v}_{rel,p}\right) &=&\sigma _{0,p}+\sigma
_{1,p}\left( \vec{v}_{rel,p}-\vec{v}_{0}\right)  \label{eqn15}
\end{eqnarray}
\begin{eqnarray}
\sigma _{H}\left( \vec{v}_{rel,H}\right) &=&\sigma _{0,H}+\sigma
_{1,H}\left( \vec{v}_{rel,H}-\vec{v}_{0}\right)  \label{eqn15aa}
\end{eqnarray}
We take as $\vec{v}_{0}$ appropriate bulk velocities $\vec{u}_{p}$, $\vec{u}_{H}$
and we change variables so that now we will be in a reference system
co-moving with the protons or vice versa with the hydrogen atoms. The
coefficients $\sigma _{0}$ and $\sigma_{1}$ in equations \ref{eqn15} and
\ref{eqn15aa} are calculated from the following formulae:
\begin{equation}
\sigma _{0,H} =\left( A+B\log \bar{u}_{H}\right)^{2}  \label{eqn15a}
\end{equation}
\begin{equation}
\sigma _{1,H} =\frac{d\sigma }{dv}\left( \bar{u}_{H}\right) =2\left(
A+B\log \bar{u}_{H}\right) \frac{B}{u_{H}}  \label{eqn15b}
\end{equation}
\begin{equation}
\sigma _{0,p} =\left( A+B\log \bar{u}_{p}\right)^{2}  \label{eqn15c}
\end{equation}
\begin{equation}
\sigma _{1,p} =\frac{d\sigma }{dv}\left( u_{p}\right) =2\left( A+B\log
\bar{u}_{p}\right) \frac{B}{u_{p}}  \label{eqn15d}
\end{equation}
where $\bar{u}_{p}$, $\bar{u}_{H}$ are velocities normalized by the velocity
of 1 cm/s. The relative velocity $\vec{v}_{rel}$ in equations \ref{eqn15}
and \ref{eqn15aa}\ is expressed as $\vec{v}_{rel,p}=\left\vert \vec{u}_{p}-
\vec{v}\right\vert $ for protons and as
$\vec{v}_{rel,H}=\left\vert \vec{u}_{H}-\vec{v}\right\vert $
for hydrogen atoms.

In the co-moving reference systems the Maxwellian core distribution
functions are the following:
\begin{eqnarray}
f_{0,p}\left( v^{\prime },s\right) &=&\frac{n_{p}\left( s\right) }{\left(
\pi ~c_{p}\left( s\right) \right) ^{3/2}}\exp \left[ -\frac{v^{\prime 2}}{
c_{p}\left( s\right) ^{2}}\right]  \label{eqn16}
\end{eqnarray}
\begin{eqnarray}
f_{0,H}\left( v^{\prime },s\right) &=&\frac{n_{H}\left( s\right) }{\left(
\pi ~c_{H}\left( s\right) \right) ^{3/2}}\exp \left[ -\frac{v^{\prime 2}}{
c_{H}\left( s\right) ^{2}}\right]  \label{eqn17}
\end{eqnarray}
where $v^{\prime }$ are specific velocities in the appropriate co-moving
frames. Evaluating equations \ref{eqn13} and \ref{eqn14} with the use of
\ref{eqn15}, \ref{eqn15aa} brings the following expressions for the coupling terms:
\begin{eqnarray}
\label{eqn18}
\frac{\delta ^{-}f_{H}}{\delta t} &=&f_{H}\left[ \sigma _{0,p}\int
v_{rel,p}~f_{0,p}~d^{3}v_{p}^{\prime }+ \right. \\
  &+& \left. \sigma _{1,p}\int
v_{rel,p}^{2}~f_{0,p}~d^{3}v_{p}^{\prime }\right] \nonumber
\end{eqnarray}
\begin{eqnarray}
\label{eqn19}
\frac{\delta ^{-}f_{p}}{\delta t} &=&f_{p}\left[ \sigma _{0,H}\int
v_{rel,H}~f_{0,H}~d^{3}v_{H}^{\prime }+ \right. \\
 & + & \left. \sigma _{1,H}\int
v_{relH}^{2}~f_{0,H}~d^{3}v_{H}^{\prime }\right]  \nonumber
\end{eqnarray}
The relative velocities under the integrands are defined as follows:
\begin{equation}
v_{rel,p}^{2} =v^{2}+u_{p}^{2}-2v~u_{p}~\cos \theta  \label{eqn20}
\end{equation}
\begin{equation}
v_{rel,H}^{2} =v^{2}+u_{H}^{2}-2v~u_{H}~\cos \theta  \label{eqn21}
\end{equation}

From Fahr \& Mueller (1967)\nocite{fahr_mueller:67} and Ripken \& Fahr
(1983)\nocite{ripken_fahr:83a}, the mean relative velocities coming out from
the first integral in equations \ref{eqn18} and \ref{eqn19} are given by
formulae:
\begin{equation}
\left\langle v_{rel,p}\right\rangle =\frac{c_{p}}{\sqrt{\pi }}\exp \left[ -
\frac{u_{p}^{2}}{c_{p}^{2}}\right] +u_{p}\left( 1+\frac{c_{p}}{2u_{p}}
\right) {\rm erf}\left( \frac{u_{p}}{c_{p}}\right)  \label{eqn23}
\end{equation}
\begin{equation}
\left\langle v_{rel,H}\right\rangle =\frac{c_{H}}{\sqrt{\pi }}\exp \left[ -
\frac{u_{H}^{2}}{c_{H}^{2}}\right] +u_{H}\left( 1+\frac{c_{H}}{2u_{H}}
\right) {\rm erf}\left( \frac{u_{H}}{c_{H}}\right)  \label{eqn24}
\end{equation}
For the second integral in formulae \ref{eqn18} and \ref{eqn19}, we
notice that the integration over $\theta $ will bring to 0 the terms
with $\cos \theta $ and that -- since $u_{p}$, $u_{H}$ are local
quantities which do not depend on $v$ -- the second term from the
relative velocity expansion (equations \ref{eqn20} and \ref{eqn21}) will
simply yield $u_{p}^{2}$ times the result of integration of the core
distribution function over the velocity space. We
therefore can denote this result simply by $ n_{0,H}$ and $n_{0,p}$,
respectively. The remaining integrations of the mean square velocity
with respect to the core distribution functions yield simply the local
thermal velocities and thus finally the integrations defined in
equations \ref{eqn18} and \ref{eqn19} lead to the following results:
\begin{equation}
\frac{\delta ^{-}f_{H}}{\delta t} =n_{p}~f_{H}~\left[ \sigma
_{0,p} \left\langle v_{rel,p}\right\rangle +\sigma _{1,p} \left(
u_{T,p}^{2}+u_{p}^{2}\right) \right]  \label{eqn25}
\end{equation}
\begin{equation}
\frac{\delta ^{-}f_{p}}{\delta t} =n_{H}~f_{p}~\left[ \sigma
_{0,H} \left\langle v_{rel,H}\right\rangle +\sigma _{1,H} \left(
u_{T,H}^{2}+u_{H}^{2}\right) \right]  \label{eqn26}
\end{equation}
with the mean relative velocities defined in Equations \ref{eqn23} and
\ref{eqn24}. The set of differential equations, originally defined by
formulae \ref{eqn1} and \ref{eqn2}, is now obtained in the following form:
\begin{eqnarray}
v\cos \theta \frac{df_{1,H}}{ds} &=&-v\cos \theta \frac{df_{0,H}}{ds}-\frac{
f_{1,H}}{\tau _{H}}- n_{p}\left( f_{0,H}+f_{1,H}\right) \times \nonumber \\
&\times& \left[ \sigma _{0,p}\left\langle v_{rel,p}\right\rangle +
\sigma _{1,p}\left( c_{p}^{2}+u_{p}^{2}\right)
\right] + \label{eqn27}\\
&+&n_{H}\left( f_{0,p}+f_{1,p}\right) \times \nonumber \\
&\times& \left[ \sigma _{0,H}\left\langle
v_{rel,H}\right\rangle +\sigma _{1,H}\left( c_{H}^{2}+u_{H}^{2}\right)
\right]  \nonumber 
\end{eqnarray}
\begin{eqnarray}
\label{eqn28}
v\cos \theta \frac{df_{1,p}}{ds} &=&-v\cos \theta \frac{df_{1,p}}{ds}-\frac{
f_{1,p}}{\tau _{p}}- n_{H}\left( f_{0,p}+f_{1,p}\right)  \times \nonumber \\
&\times& \left[ \sigma _{0,H}\left\langle
v_{rel,H}\right\rangle +\sigma _{1,H}\left( c_{H}^{2}+u_{H}^{2}\right)
\right] + \\
&+&n_{p}\left( f_{0,H}+f_{1,H}\right) \times \nonumber \\
&\times& \left[ \sigma _{0,p}\left\langle
v_{rel,p}\right\rangle +\sigma _{1,p}\left( c_{p}^{2}+u_{p}^{2}\right)
\right]  \nonumber
\end{eqnarray}
The above equations can be rewritten so that the terms multiplied by
$f_{1,p;H}$ are collected together:
\begin{eqnarray}
\label{eqn27a}
v\cos \theta \frac{df_{1,H}}{ds} &=&-v\cos \theta
~\frac{df_{0,H}}{ds}-\frac{f_{1,H}}{\tau_H} - \\
&-&f_{1,H} n_{p}\,\left( \sigma _{0,p}\left\langle
v_{rel,p}\right\rangle +\sigma _{1,p}\left( c_{p}^{2}+u_{p}^{2}\right)
\right) +  \nonumber \\
&+&f_{1,p}~n_{H}\left[ \sigma _{0,H}\left\langle v_{rel,H}\right\rangle
+\sigma _{1,H}\left( c_{H}^{2}+u_{H}^{2}\right) \right] -  \nonumber \\
&-&f_{0,H}~n_{p}\left[ \sigma _{0,p}\left\langle v_{rel,p}\right\rangle
+\sigma _{1,p}\left( c_{p}^{2}+u_{p}^{2}\right) \right] +  \nonumber \\
&+&f_{0,p}~n_{H}\left[ \sigma _{0,H}\left\langle v_{rel,H}\right\rangle
+\sigma _{1,H}\left( c_{H}^{2}+u_{H}^{2}\right) \right]   \nonumber
\end{eqnarray}
\begin{eqnarray}
\label{eqn28a}
v\cos \theta \frac{df_{1,p}}{ds} &=&-v\cos \theta ~\frac{df_{0,p}}{ds}
 - \frac{f_{1,p}}{\tau_p} - \\
&-&f_{1,p}n_{H}\left( \sigma _{0,H}\left\langle
v_{rel,H}\right\rangle +\sigma _{1,H}\left( c_{H}^{2}+u_{H}^{2}\right)
\right) +  \nonumber \\
&+&f_{1,H}~n_{p}\left[ \sigma _{0,p}\left\langle v_{rel,p}\right\rangle
+\sigma _{1,p}\left( c_{p}^{2}+u_{p}^{2}\right) \right] -  \nonumber \\
&-&f_{0,p}~n_{H}\left[ \sigma _{0,H}\left\langle v_{rel,H}\right\rangle
+\sigma _{1,H}\left( c_{H}^{2}+u_{H}^{2}\right) \right] +  \nonumber \\
&+&f_{0,H}~n_{p}\left[ \sigma _{0,p}\left\langle v_{rel,p}\right\rangle
+\sigma _{1,p}\left( c_{p}^{2}+u_{p}^{2}\right) \right]  \nonumber
\end{eqnarray}
These equations hence formally represent the following linear system of
differential equations:
\begin{eqnarray}
v\cos \theta \frac{df_{1,H}}{ds} &=&A\left( s\right) +B\left( s\right)
~f_{1,H}+C\left( s\right) ~f_{1,p}  \label{eqn27b}
\end{eqnarray}
\begin{eqnarray}
v\cos \theta \frac{df_{1,p}}{ds} &=&D\left( s\right) +E\left( s\right)
~f_{1,H}+F\left( s\right) ~f_{1,p}  \label{eqn28b}
\end{eqnarray}
with the multiplicative factors $A,B,C,D,E,F$ defined as follows:
\begin{eqnarray}
\label{eqn275a}
A\left( s\right) &=&-v\cos \theta ~\frac{df_{0,H}}{ds}-  \\
&&-f_{0,H}~n_{p}\left[ \sigma _{0,p}\left\langle v_{rel,p}\right\rangle
+\sigma _{1,p}\left( c_{p}^{2}+u_{p}^{2}\right) \right] +  \nonumber \\
&&+f_{0,p}~n_{H}\left[ \sigma _{0,H}\left\langle v_{rel,H}\right\rangle
+\sigma _{1,H}\left( c_{H}^{2}+u_{H}^{2}\right) \right]  \nonumber
\end{eqnarray}
\begin{eqnarray}
\label{eqn275b}
B\left( s\right) &=&-\left[ \frac{1}{\tau _{H}}+n_{p}\left( \sigma
_{0,p}\left\langle v_{rel,p}\right\rangle +\sigma _{1,p}\left(
c_{p}^{2}+u_{p}^{2}\right) \right) \right]
\end{eqnarray}
\begin{eqnarray}
\label{eqn275c}
C\left( s\right) &=&n_{H}\left[ \sigma _{0,H}\left\langle
v_{rel,H}\right\rangle +\sigma _{1,H}\left( c_{H}^{2}+u_{H}^{2}\right)
\right]
\end{eqnarray}
\begin{eqnarray}
\label{eqn285a}
D\left( s\right) &=&-v\cos \theta ~\frac{df_{0,p}}{ds}+  \\
&&+f_{0,H}~n_{p}\left[ \sigma _{0,p}\left\langle v_{rel,p}\right\rangle
+\sigma _{1,p}\left( c_{p}^{2}+u_{p}^{2}\right) \right] -  \nonumber \\
&&-f_{0,p}~n_{H}\left[ \sigma _{0,H}\left\langle v_{rel,H}\right\rangle
+\sigma _{1,H}\left( c_{H}^{2}+u_{H}^{2}\right) \right]  \nonumber
\end{eqnarray}
\begin{eqnarray}
\label{eqn285b}
E\left( s\right) &=&n_{p}\left[ \sigma _{0,p}\left\langle
v_{rel,p}\right\rangle +\sigma _{1,p}\left( c_{p}^{2}+u_{p}^{2}\right)
\right]
\end{eqnarray}
\begin{eqnarray}
\label{eqn285c}
F\left( s\right) &=&- \frac{1}{\tau _{p}}-n_{H}\left( \sigma
_{0,H}\left\langle v_{rel,H}\right\rangle +\sigma _{1,H}\left(
c_{H}^{2}+u_{H}^{2}\right) \right)
\end{eqnarray}

The derivatives of the unperturbed distribution functions given by Equation
\ref{eqn7} and \ref{eqn8}\ are calculated as follows:
\begin{equation}
\frac{df_{0,H}}{ds}=\frac{\partial f_{0,H}}{\partial n_{H}}~\frac{dn_{H}}{ds}
+\frac{\partial f_{0,H}}{\partial u_{H}}~\frac{du_{H}}{ds}+\frac{\partial
f_{0,H}}{\partial c_{H}}~\frac{dc_{H}}{ds}  \label{eqn29}
\end{equation}
and analogously for the function $f_{0,p}$. The terms of equation \ref{eqn29}
are evaluated as follows:
\begin{equation}
\frac{\partial f_{0,H}}{\partial n_{H}}=\frac{f_{0,H}}{n_{H}}  \label{eqn30}
\end{equation}
\begin{equation}
\frac{\partial f_{0,H}}{\partial u_{H}}=-2f_{0,H}\frac{u_{H}-v~\cos \theta }{
c_{H}^{2}}  \label{eqn31}
\end{equation}
\begin{equation}
\frac{\partial f_{0,H}}{\partial c_{H}}=\frac{f_{0,H}}{c_{H}}\left( 2\frac{
v^{2}+u_{H}^{2}-2v~u_{H}~\cos \theta }{c_{H}^{2}}-\frac{3}{2}\right)
\label{eqn32}
\end{equation}
Collecting them all together we obtain the following expression for the
derivatives of the unperturbed distribution functions:
\begin{eqnarray}
\label{eqn33}
\frac{df_{0,H}}{ds} &=&f_{0,H}\left[ \frac{\frac{dn_{H}}{ds}}{n_{H}}-\frac{2}
{c_{H}^{2}}\left( \left( u_{H}-v\cos \theta \right) \frac{du_{H}}{ds} - \right.
\right. \\
 &-& \left. \left.\left( \frac{v^{2}+u_{H}^{2}-2vu_{H}~\cos \theta }{c_{H}}-\frac{3}{4}
\right) \frac{dc_{H}}{ds}\right) \right] \nonumber
\end{eqnarray}
\begin{eqnarray}
\label{eqn34}
\frac{df_{0,p}}{ds} &=&f_{0,p}\left[ \frac{\frac{dn_{p}}{ds}}{n_{p}}-\frac{2}
{c_{p}^{2}}\left( \left( u_{p}-v~\cos \theta \right) \frac{du_{p}}{ds} - \right.
\right. \\
 &-& \left. \left. \left( \frac{v^{2}+u_{p}^{2}-2v~u_{p}~\cos \theta }{c_{p}}-\frac{3}{4}
\right) \frac{dc_{p}}{ds}\right) \right] \nonumber
\end{eqnarray}

Now we need the relaxation times, used in equations \ref{eqn3} and \ref{eqn4}.
For the relaxation of the neutral hydrogen gas, the only mechanism is
elastic H -- H collisions. The corresponding relaxation time is numerically given by
Brinkmann (1970)\nocite{brinkmann:70} and Fahr (1996)\nocite{fahr:96} in the following form:
\begin{equation}
\tau _{H}=\frac{6.62\cdot 10^{10}}{n_{H}\sqrt{T_{H}}}=\frac{6.62\cdot 10^{10}
}{n_{H}~c_{H}}\sqrt{\frac{2K}{m}}=\frac{8.51\cdot 10^{14}}{n_{H}~c_{H}}~
{\rm s}  \label{eqn35}
\end{equation}
for $n_H$ expressed in cm$^{-3}$ and $c_H$ in cm/s. The period $\tau_H$
turns out to be very long in comparison to charge exchange periods $\tau_{ex}$,
i.e., $\tau_H/\tau_{ex} \simeq 10$. The relaxation time
for protons is more difficult to calculate. If, however,
the secondary protons have a bulk velocity small enough to not be 
shifted outside the broad thermal envelope of the
``hydrodynamic'' proton population (in other words, the net
distribution function does not feature two distinct peaks), then the
wave-particle interaction by a two-stream instability should not be
significant. If, in addition, the pre-existing turbulence level in the
unperturbed LIC\ is small at the distance scale comparable to and smaller
than the size of the interface, then also non-linear interaction with MHD
waves (mainly Alfv\'{e}n waves) may be negligible and the only relaxation
mechanism will be Coulomb collisions between the old and newly-created
protons. A more in-depth discussion will be presented in Sec.4 at the end of
this paper.
\begin{figure*}
\centering
\includegraphics[width=17cm]{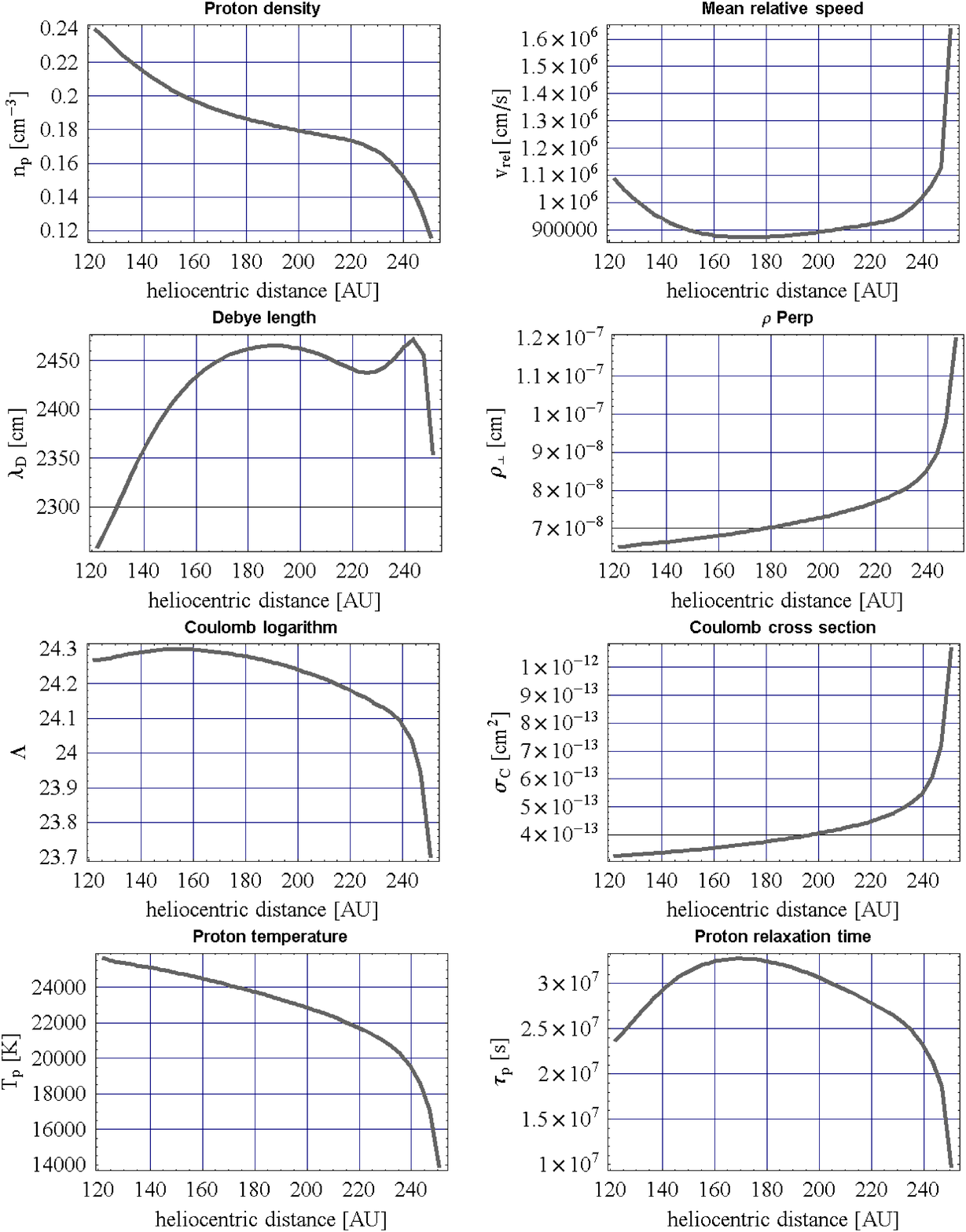}
\caption{Parameters relevant to compute the proton relaxation time, discussed
in Eq. \ref{eqn39} through \ref{eqn38}, as functions of the heliocentric distance
between the bowshock and the heliopause, based on the parameters shown in
Fig. \ref{figPParam}.}
\label{figprotonRelax}
\end{figure*}

The proton relaxation time is defined as the time between Coulomb collisions
of a proton traveling at $v$ and protons belonging to the core
population, described by the proton distribution function $f_{0}$ being a
Maxwellian shifted by $u_{p}$ with thermal velocity $c_{p}$, and is obtained
from the mean free path with respect to Coulomb collisions $\lambda_C$ in the form:
\begin{equation}
\tau _{p}=\frac{\lambda _{C}}{v_{rel,p}}=\frac{1}{\frac{1}{2}n_{p}~\sigma
_{C}~\left\langle v_{rel,p}\right\rangle }  \label{eqn39}
\end{equation}
where $\frac{1}{2}$ is the energy transfer efficiency term, equal to the
reduced mass of the colliding particles: $2\,m_{p}~m_{p}/\left(
m_{p}+m_{p}\right)^{2}$.

The angle-averaged mean momentum-transfer Coulomb collision cross section $\sigma_C$
in Eq. \ref{eqn39} is given by the textbook formula
(e.g., Oraevskij 1989\nocite{oraevskij:89}):
\begin{equation}
\sigma _{C}=\sigma _{C,\perp }\Lambda   \label{eqn36}
\end{equation}%
where $\Lambda $ is the Coulomb logarithm defined as
\begin{equation}
\Lambda =\ln \frac{\lambda _{D}}{\rho _{\perp }}  \label{eqn36a}
\end{equation}
$\sigma _{C,\perp }$ is defined by the formula:
\begin{equation}
\sigma _{C,\perp }=\pi \rho _{\perp }^{2}  \label{eqn36b}
\end{equation}
and $\rho _{\perp }$ in Eq.\ref{eqn36b} is defined by
\begin{equation}
\frac{e^{2}}{\rho _{\perp }}=K~T_{p}  \label{eqn36c}
\end{equation}
thus yielding
\begin{equation}
\rho _{\perp }=\frac{e^{2}}{K~T_{p}}  \label{eqn36d}
\end{equation}
so consequently
\begin{equation}
\sigma _{C,\perp }=\frac{\pi e^{4}}{\left( K~T_{p}\right)^{2}}
\label{eqn36e}
\end{equation}
In Eqs \ref{eqn36} through \ref{eqn36e} $K$ is Boltzmann constant;
$\left\langle v_{rel,p}\right\rangle $ is the mean relative velocity of the
test particle (the newly-created proton) with respect to the unperturbed
proton population described by $f_{0,p}$, defined in Eq.~\ref{eqn23}; and $e$
is the cgs elementary charge. Deep inside the interface, at 200~AU from the
Sun, the proton density $n_{p}=0.18$ cm$^{-3}$, proton temperature
$T_{p}=2.3~10^{4}$ K, and the relative speed $\left\langle
v_{rel,p}\right\rangle =9~10^{5}$ cm/s. In Eq.\ref{eqn36a} $\lambda _{D}$ is
the Debye length, defined by the formula:
\begin{equation}
\lambda _{D}=\sqrt{\frac{KT_{p}}{4\pi ~n_{p}~e^{2}}}=c_{p}\sqrt{\frac{m}
{8\pi ~n_{p}~e^{2}}}=6.9\sqrt{\frac{T_{p}}{n_{p}}}  \label{eqn37}
\end{equation}
which for the above-mentioned conditions in the heliospheric interface is
equal to $\sim $2460~cm, and the Coulomb logarithm $\Lambda $, defined in Eq.
\ref{eqn36a}, to $\sim $24.25. Thus the total Coulomb cross section $\sigma
_{C}$, defined in Eq.\ref{eqn36}, is equal to $\sim 4~10^{-13}$~cm$^{2}$.
Hence the proton relaxation time within the outer interface, at $\sim 200$ AU from
the Sun, is of the order of a year:
\begin{equation}
\tau _{p}\simeq 3.1~10^{7}{\rm s}  \label{eqn38}
\end{equation}
These times $\tau_p$ become even much larger in the inner interface and in the heliotail.
These estimates may express the fact that relaxation processes are fairly
unimportant here as well for the protons as especially for the H-atoms
because the relaxation periods are comparable to the mean passage times or
structuring times
\begin{equation}
\tau _{s}=\frac{n_{H,p}}{u_{H,p}\frac{dn_{H,p}}{ds}}\approx 10^{8}{\rm s}
\label{eqn38a}
\end{equation}
In the final section of this paper we shall, however, come back to this
point. The dependence of all the above referenced quantities on the axial
coordinates counted along the axis in the interface is shown in Figs
\ref{figprotonRelax}.

\section{Results and discussion}

\begin{figure*}
\centering
\includegraphics[width=17cm]{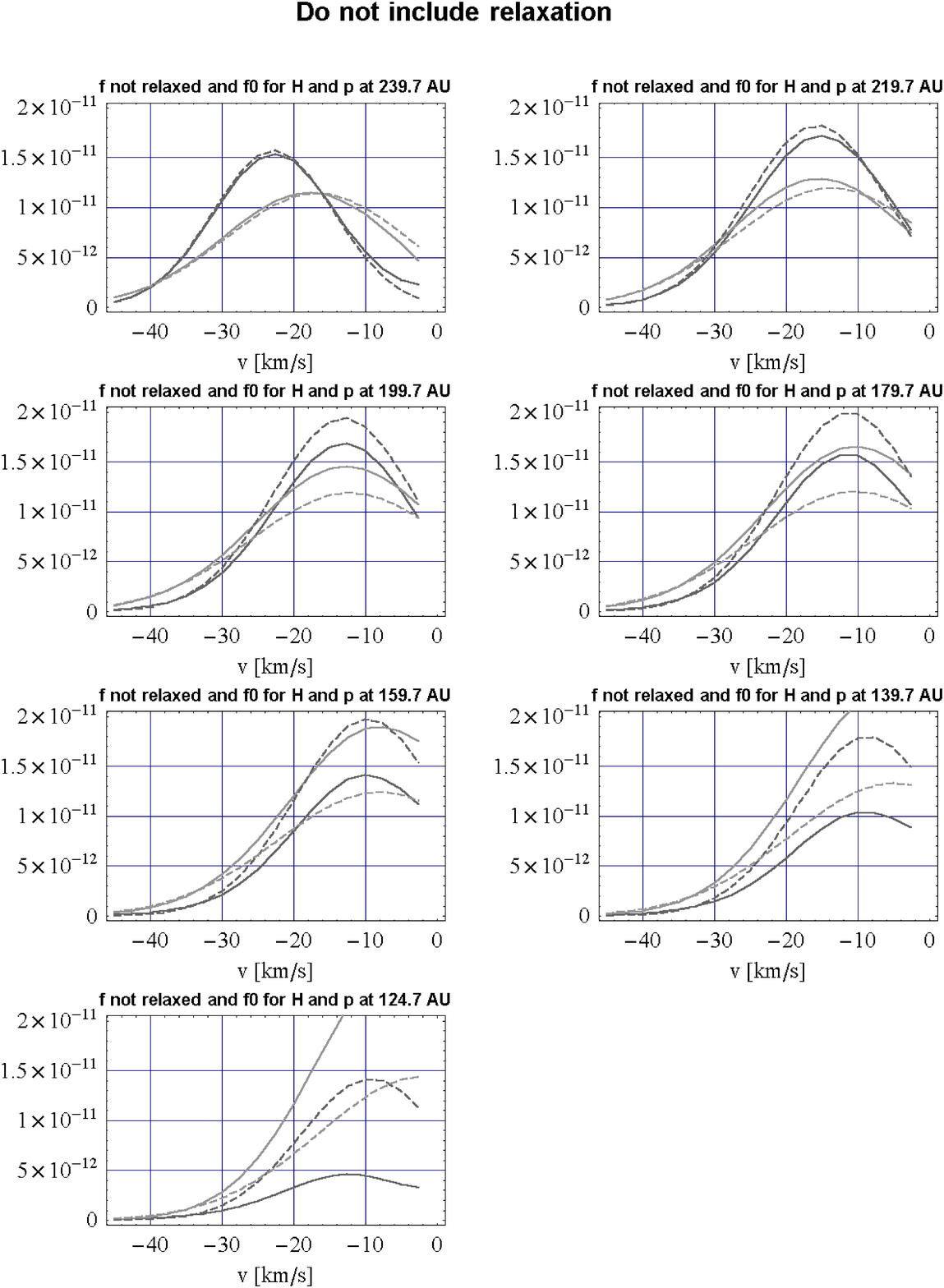}
\caption{Hydrogen and proton distribution functions: no proton relaxation case.
Light gray lines correspond to protons and dark gray to H-atoms. Shown at
selected distances from the Sun (in the region between the bow shock and the
heliopause) are the unperturbed Maxwellian distribution functions, drawn with
broken lines, and the net distribution functions $f_{p;H}=f_{0,p;H}+f_{1,p;H}$,
drawn with the solid lines and expressed in atoms~cm$^{-6}$~s$^{-3}$.}
\label{figSoluNoRelax}
\end{figure*}

\begin{figure*}
\centering
\includegraphics[width=17cm]{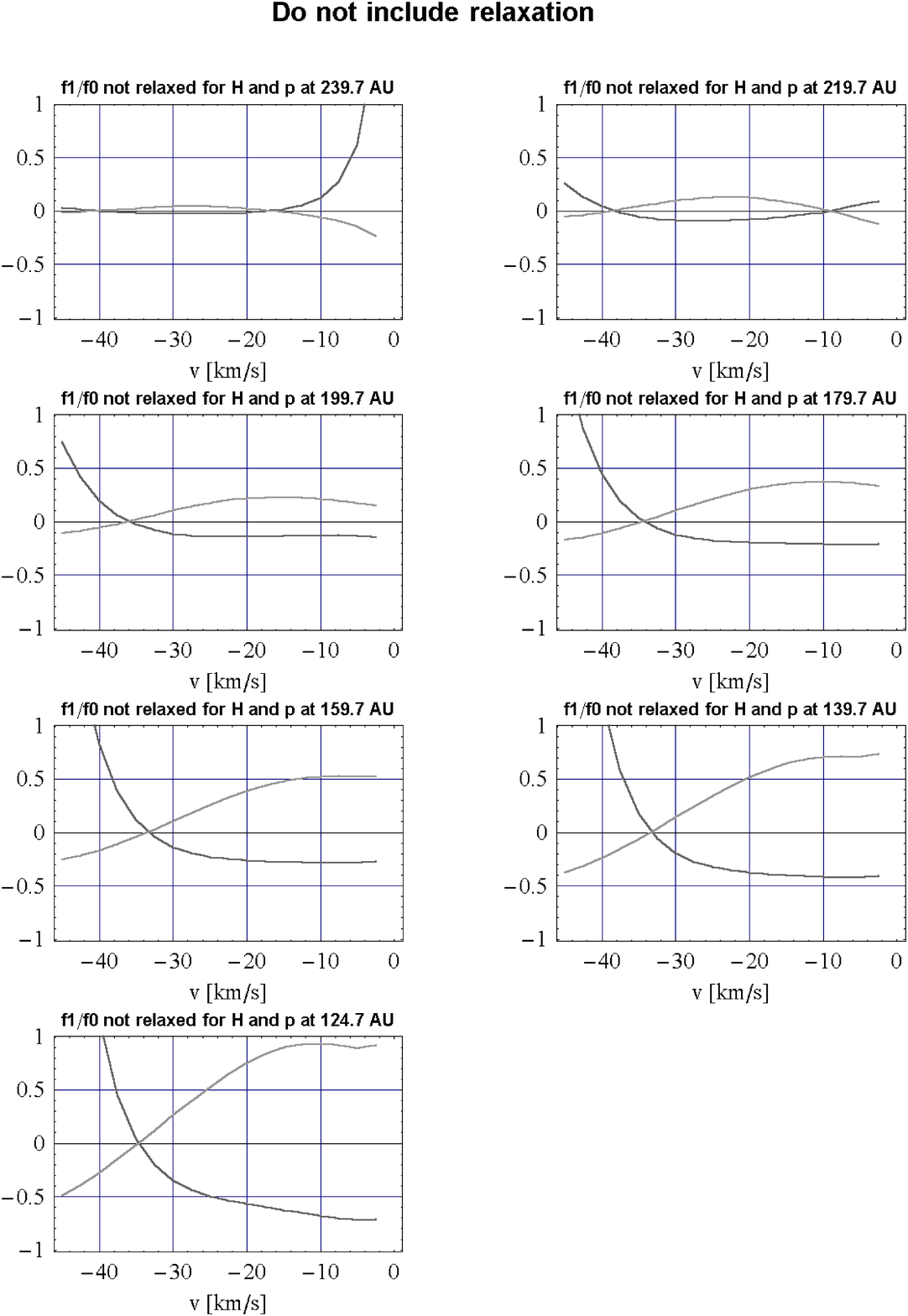}
\caption{Ratios of the perturbation functions $f_{1,p;H}$ to the corresponding
unperturbed distribution functions $f_{0,p;H}$: the $f_1/f_0$ ratios, in the
case of no proton relaxation operating, drawn for the H-atoms in dark gray and
for the protons in light gray.}
\label{figSoluNoRelaxRel}
\end{figure*}

\begin{figure*}
\centering
\includegraphics[width=17cm]{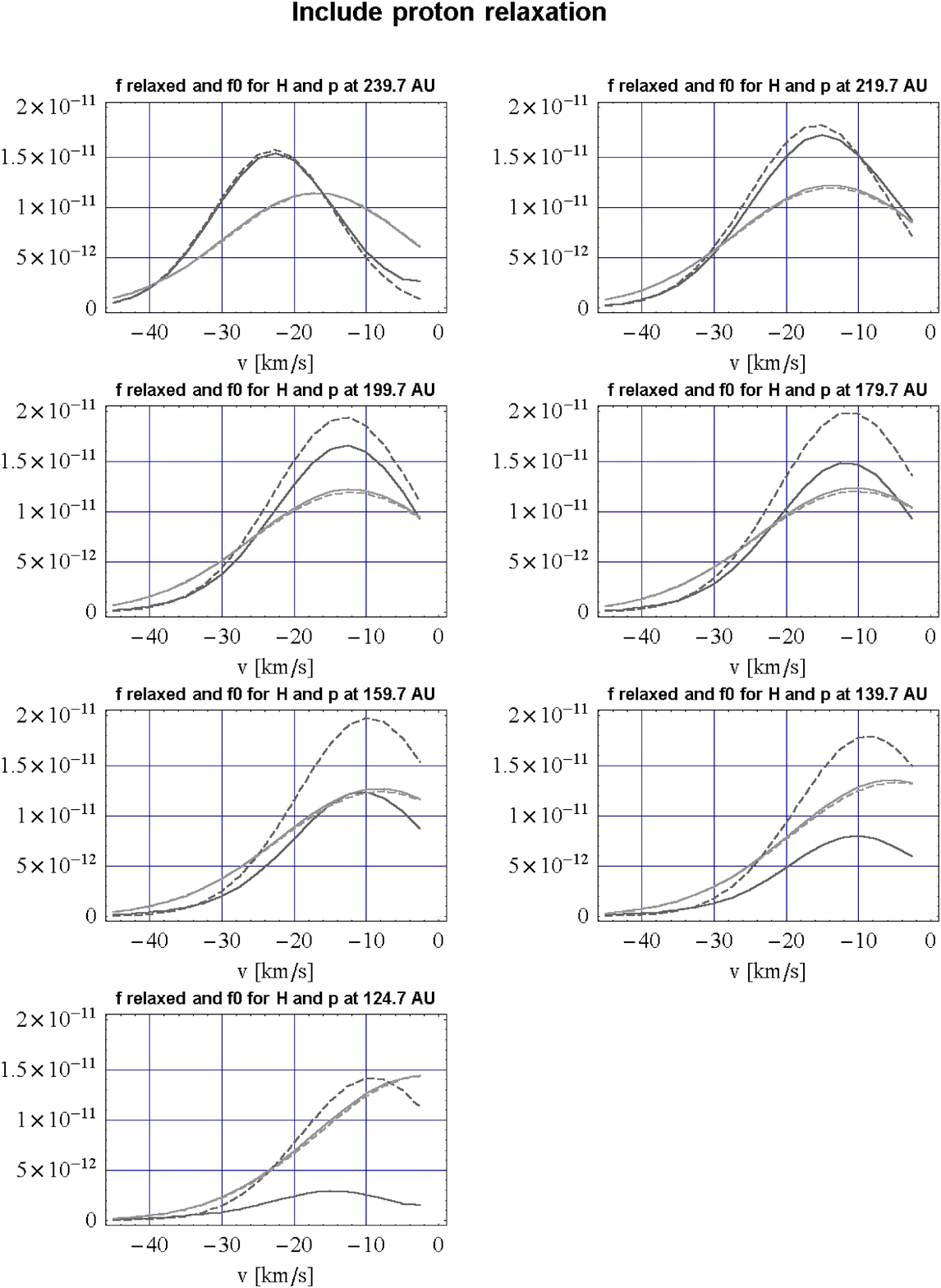}
\caption{Hydrogen and proton distribution functions: proton Coulomb relaxation case.
Light gray lines correspond to protons and dark gray to H-atoms. Shown at
selected distances from the Sun (in the region between the bow shock and the
heliopause) are the unperturbed Maxwellian distribution functions, drawn with
broken lines, and the net distribution functions $f_{p;H}=f_{0,p;H}+f_{1,p;H}$,
drawn with the solid lines and expressed in atoms~cm$^{-6}$~s$^{-3}$.}
\label{figSoluRelax}
\end{figure*}

\begin{figure*}
\centering
\includegraphics[width=17cm]{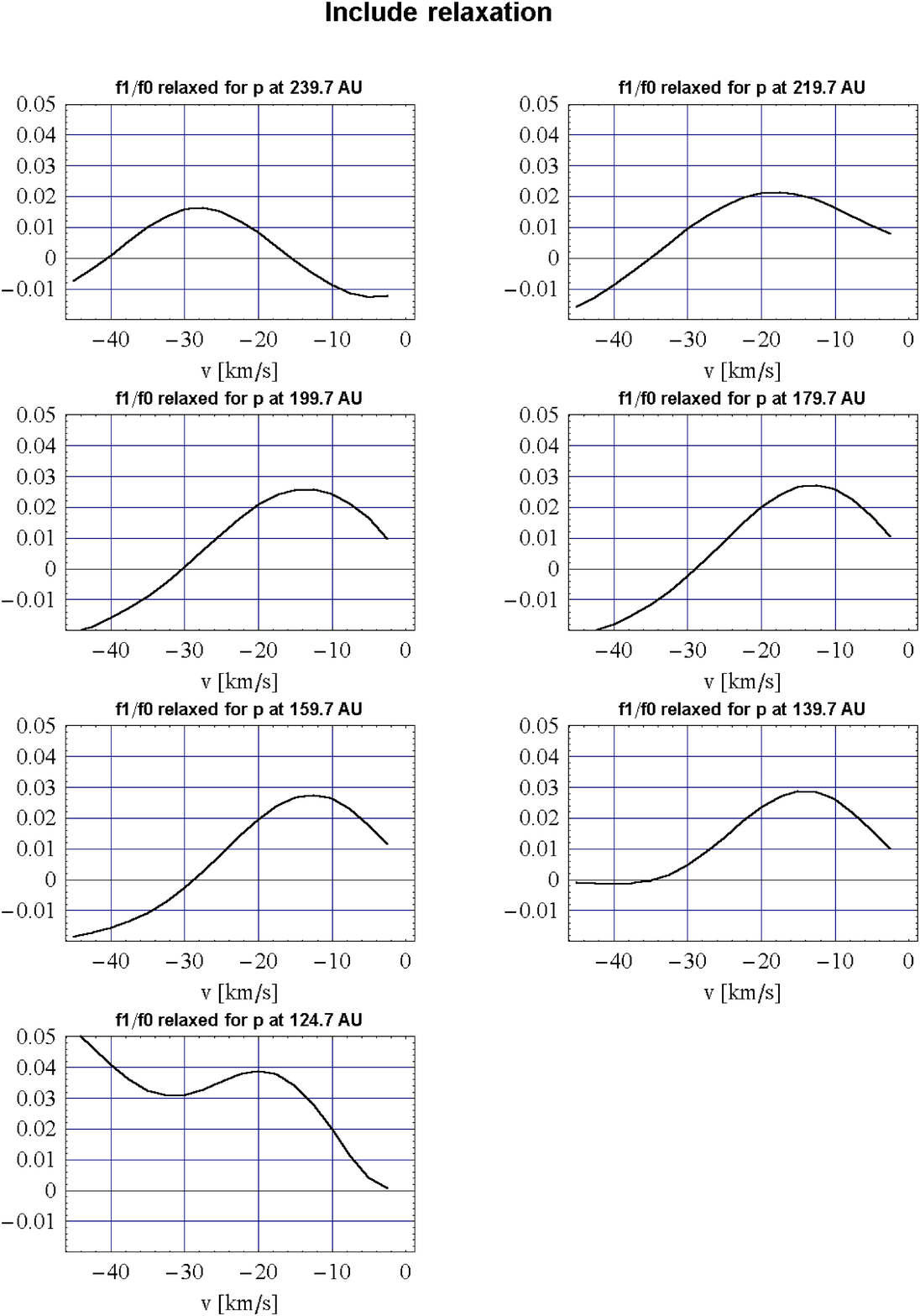}
\caption{Ratios of the proton perturbation function to the proton unperturbed
distribution function: $f_1/f_0$ for the proton Coulomb relaxation operating.}
\label{figSoluRelaxRel}
\end{figure*}

We solved the system of equations \ref{eqn27b}, \ref{eqn28b}
(separarately for each $v$) using the hydrodynamic core distribution
functions given in Eqs \ref{eqn7} and \ref{eqn8}, with the parameters
obtained from hydrodynamic simulations by Fahr et al.
(2000)\nocite{fahr_etal:00} and shown in Figures \ref{figHParam} and
\ref{figPParam}. In the approach presented here, where H-atoms are
represented by a mono-Maxwellian plus kinetic deviations, we can only
represent solutions of the distribution function for positive values of
H-atom velocities. In order to present solutions for negative values of
H-atom velocities we would need to integrate Equs \ref{eqn1} and
\ref{eqn2} in the opposite direction (negative increments of $s$!)
starting from an inner boundary value for the H-atom distribution
function $f_{0,H}\left(s_0, -v_H \right)$. Since our hydrodynamic H-atom
core functions vanish for negative velocity values, we cannot use any
reasonable boundary condition to calculate the H-atom distribution
functions for negative velocity values unless we admit additional H-atom
fluids, as practised by Zank et al. (1996)\nocite{zank_etal:96a}. The
results for densities $n_{p,H}$, bulk velocities $u_{p,H}$, and
temperatures $T_{p,H}$ have been calculated on the basis of the
assumption that the underlying distribution functions $f_{p,H}$ are
three-moment functions, i.e., shifted Maxwellians. In contrast, in the
kinetic theory presented in this paper we have calculated those
distribution functions actually occurring in the outer interface under
the action of charge exchange processes between the locally
colliding H-atoms and protons. These results we show in Figures
\ref{figSoluNoRelax} through \ref{figSoluRelaxRel}. In these figures we
have displayed at several interface positions on the symmetry axis
one-dimensional cuts through the kinetically resulting distribution
functions $f_{p,H}$, showing both the unperturbed hydrodynamical
distribution functions $f_{0(p,H)}$ and the kinetically perturbed,
actually resulting distribution functions
$f_{(p,H)}=f_{0(p,H)}+f_{1(p,H)}$.

First in Fig. \ref{figSoluNoRelax} we display $v_{z}$-cuts through the
distribution functions $f_{H}$ and $f_{p}$ resulting at various
positions $s=s_{i}$ in the interface, when proton-proton relaxation
processes are suppressed, but atom-atom relaxation processes are
working. As one can clearly see, pronounced deviations of both resulting
distribution functions $f_{(p,H)}$ from simple shifted Maxwellians,
derived by hydrodynamical studies, can be recognized. These are even
better identifiable in Fig. \ref{figSoluNoRelaxRel}, where we have
displayed the ratios $f_{1(p,H)}/f_{0(p,H)}$. Switching on the
proton-proton Coulomb collision relaxation, as developed in Eq.
\ref{eqn39} and as shown in Figs \ref{figSoluRelax} and
\ref{figSoluRelaxRel}, reduces these deviations in the case of proton
distribution functions to much milder degrees. Thus, under LISM
conditions as used here in this paper, the hydrodynamically derived
distribution functions for the interface protons may be considered as
good approximations, whereas the hydrodynamic approach to the H-atom
distribution definitely is a poor approximation. Thus, due to rapid
relaxation processes under the conditions adopted in this paper
deviations of protons from hydrodynamically derived Maxwellians can be
considered as fairly small. 

In contrast, however, if interfaces for different LISM conditions need
to be modelled, a kinetic revision of such proton distribution functions
becomes necessary. For instance, this would be the case for lower LISM
proton densities, because the proton-proton relaxation rate/period
drastically goes down/up. For example, a LISM proton density of
$n_{p,\infty }=0.01$ cm$^{-3}$ would increase the relaxation rate by
more than a factor of 10 with respect to the one used here, thus meaning
in consequence that non-relaxated proton distribution functions must be
expected. This would become important in the inner heliosheath and in
the heliotail. This all the more will be true if magnetic fields control
the interface MHD dynamics and by CGL-effects (Chew et al.
1956\nocite{chew_etal:56}) may lead to pronounced temperature
anisotropies.

As one can notice in Figs \ref{figSoluRelax} and \ref{figSoluRelaxRel},
the closer one approaches the Sun starting from the outer bow shock, the
more the resulting kinetic distribution function $f_{(p,H)}$ deviates
from the unperturbed hydrodynamic distribution $f_{0(p,H)}$, even
reaching nonlinear amounts of such deviations. The specific form of
these deviations thereby are not systematic with the coordinate $s$ and
could only perhaps be studied in more detail by the analysis of the
higher moments of the distribution function $f_{(p,H)}$, like the heat
conduction flow or the temperature anisotropies. Qualitatively at least
one can say -- as it was already anticipated in the qualitative results
sketched in Fig. \ref{figDemo} -- that the solar/antisolar wings of the
distributions $f_{(p,H)}$ are enhanced with respect to the unperturbed
ones $f_{0(p,H)}$, wherever the local bulk velocities $u_{(p,H)}$ are
larger than those $u_{(p,H)}$. Under these conditions heat conduction
flows result from the distribution functions $f_{(p,H)}$, which for
protons are directed in the direction of $\vec{u}_{H}-\vec{u}_{p}$ and
for H-atoms in the direction $\vec{u}_{p}-\vec{u}_{H}$.

This fact has two consequences for the modelling of the H-atom presence
in the inner and outer heliosphere. One is that the distribution
function $f_{H} $, which describes the injection of the LISM H-atoms
into the inner heliosphere, is different from what was assumed up to
now. This thus will influence the predictions made with respect to the
H-atom phase-space distribution in the inner heliosphere and the
related predictions on the resulting spectral and integrated
Lyman-$\alpha$ resonance glow (see, e.g., Scherer et al.
1999\nocite{scherer_etal:99}). Another consequence may be that
calculations of the absorption spectra caused by hydrogen atoms in the
hydrogen wall ahead of the heliopause (see papers by Linsky \& Wood
1996\nocite{linsky_wood:96}; Wood \& Linsky 1997\nocite{wood_linsky:97};
Gayley et al. 1997\nocite{gayley_etal:97}; Izmodenov et al.
1999\nocite{izmodenov_etal:99b}) may need revision. For instance, the
asymmetric shapes of the resulting H-atom distribution function
presented in this paper lead to asymmetric spectral absorption features
at Lyman-$\alpha $ wavelengths which could remedy the problems in the
correct representation of the observed stellar spectra (see, e.g.,
Izmodenov et al. 1999\nocite{izmodenov_etal:99b}).

\section{Outlook: Competing proton relaxation processes in heliospheric
interface}

In the above calculations we have made use of the assumption that
relaxation of the actually formed proton distribution function towards
an associated equilibrium distribution $f_{0}$ with $\delta \left( \int
f_{0}~\ln f_{0}~ d^{3}v\right)/\delta t=0$ operates on time scales
comparable to charge exchange time scales and the typical interface
passage times. As proton-proton relaxation processes we only have taken
into account p-p Coulomb interaction. This assumption should
be justified and for that purpose we are looking for some alternative
processes which could support the relaxation of the proton distribution
function.

Since the proton distribution function resulting in the interface is a
consequence of charge-exchange induced interaction between the
distributions of protons and H-atoms, the proton distribution function
$f_{p}$ approximately represents a superposition of two shifted
Maxwellians with different partial densities $ n_{0}$ and $n_{1}$, the
latter being due to ionized H-atoms. Under such circumstances it is
suggestive to first investigate whether excitation of electron plasma
waves by the positive slope of the composite distribution function may
be an important contribution to a relaxation of such non-equilibrium
distributions towards a best-adapted mono-Maxwellian.

Taking the approach presented by Davidson et al. (1970)\nocite{davidson_etal:70} 
and going to the limit of full equilibration of
electron thermal and differential proton bulk energies, one obtains a
friction force $R_{e,i}$ per unit volume acting between protons and
electrons given by (see Fahr \& Neutsch 1983\nocite{fahr_neutsch:83a}):
\begin{equation}
R_{e,i}=J\gamma _{\max }W_{k0}\sqrt{\frac{m_{e}}{KT_{e}}}  \label{eqn40}
\end{equation}
where $J=39.9$ is found as a constant, $m_{e}$ and $T_{e}$ are the electron
mass and temperature, respectively, $\gamma _{\max }$ is the maximum growth
rate of the excited plasma waves (i.e., most unstable electrostatic mode)
given by:
\begin{equation}
\gamma _{\max }=\omega _{p}\frac{\sqrt{3}}{2}\sqrt[3]{\frac{m_{e}}{4m_{p}}}
\label{eqn41}
\end{equation}
where $\omega _{p}$ is the electron plasma frequency, and $W_{k0}$ is the
total wave power which according to Davidson et al. (1970)
\nocite{davidson_etal:70} in the saturation phase attains the following value:
\begin{equation}
W_{k0}\simeq n_{e}m_{e}(u_{H}-u_{p})^{2}  \label{eqn41a}
\end{equation}
where $u_{H}$ and $u_{p}$ are the z-components of the local bulk velocities
of the H-atoms and the protons, respectively. The above mentioned friction
force $R_{e,i}$ operates with the tendency to reduce the differential bulk
kinetic energy in the proton distribution function. Thus one can derive a
typical relaxation time $\tau _{e,i}$ with the following relation:
\begin{equation}
\ \tau _{e,i}\simeq \frac{n_{1}(\frac{1}{2}m_{p}(u_{H}-u_{p})^{2})}{
R_{e,i}u_{p}}  \label{eqn42}
\end{equation}%
Inserting values like $n_{0}=0.1$~cm$^{-3}$, $n_{1}/n_{0}=0.1$, $u_{H}=25$~km/s,
$u_{p}=10$~km/s, and $KT_{e}=\frac{1}{4}m_{e}(u_{H}-u_{p})^{2}$ one gets the result: $\tau _{e,i}\simeq 2000$~s.

From the above estimate one should expect to have the relaxation of the
non-equilibrium proton distribution occurring via electron plasma wave
excitations within astonishingly short times $\tau_{e,i}$ of the order of
one or a few hours.

One should, however, check the validity of the assumptions made in the
above derivation of the time $\tau _{e,i}$. As it turns out, the results
of Davidson et al. (1970) \nocite{davidson_etal:70} can in principle
be adopted only if the two Maxwellian peaks in the superposed
distribution function are clearly separated from each other and do not
essentially overlap by their Maxwellian wings. When looking at the
conditions which in fact are prevailing at least on the interstellar
side of the heliospheric interface it becomes fairly evident that this
condition is by far not fulfilled, since the proton thermal velocity is
of the order of 20~km/s and the separation of the bulks only amounts to
perhaps 15~km/s.

This conclusion can also be supported by looking into the Landau damping
rates or the plasma oscillation growth rates $\gamma _{p}(\omega ,k)$ which
are connected to the derivatives with respect to $v_{z}$ of the normalized
distribution function $f_{p,z}^{\times }=\int \int f_{p}^{\times}dv_{x}dv_{y}$ 
by the following relation:
\begin{equation}
\gamma _{p}(\omega ,k)\simeq \frac{\omega _{p}^{3}}{\pi k^{2}}\left\vert 
\frac{\partial f_{p,z}^{\times }}{\partial v_{z}}\right\vert _{v_{z}=
\omega/k}=\omega _{p}\frac{v_{z}^{2}}{\pi }\left\vert \frac{\partial
f_{p,z}^{\times }}{\partial v_{z}}\right\vert _{v_{z}}  \label{eqn43}
\end{equation}
where the normalized distribution function $f_{p,z}^{\times }$ may
approximately be represented by:
\begin{eqnarray}
f_{p,z}^{\times }(w_{z})&\simeq& \frac{1}{\sqrt{\pi }} \exp 
\left[ -w_{z}^{2}\right] +\frac{1}{\sqrt{\pi }}\frac{n_{1}}{n_{0}}\left( \frac{T_{p}}{T_{H}}%
\right)^{1/2} \times \\ 
& \times& \exp \left[ -\frac{T_{p}}{T_{H}}\left(
W_{H}-W_{p}-w_{z}\right) ^{2}\right]  \nonumber \label{eqn44}
\end{eqnarray}
Here the following denotations were used: 
$w_{z}^{2}=m_{p}v_{z}^{2}/(2KT_{p}) $; $W_{H}^{2}=m_{p}u_{H}^{2}/(2KT_{p})$;
\ $W_{p}^{2}=m_{p}u_{p}^{2}/(2KT_{p})$; $T_{p}$ and $T_{H}$ are the
temperatures of the bulk of the protons and of the H-atoms, respectively.
Now from the above relations one finds:
\begin{eqnarray}
\frac{\partial f_{p,z}^{\times }}{\partial w_{z}}&\simeq& \frac{-2w_{z}}{\sqrt{
\pi }}\exp [-w_{z}^{2}]+\frac{2}{\sqrt{\pi }}\frac{n_{1}}{n_{0}}\left( \frac{
T_{p}}{T_{H}}\right) ^{3/2} \times \\
 &\times& (W_{H}-W_{p}-w_{z})\exp \left[ -\frac{T_{p}}{T_{H}
}\left( W_{H}-W_{p}-w_{z}^{2}\right) \right] \nonumber \label{eqn45}
\end{eqnarray}

The most positive contribution comes from the second hump in the
distribution function to $f_{p,z}^{\times }(w_{z})$ and leads to positive
growth rates evidently coming from regions of $w_{z}$ where the following
relation is valid: $W_{H}-W_{p}-w_{z}\simeq 1$, which then leads to:
\begin{eqnarray}
\left\vert \frac{\partial f_{p,z}^{\times }}{\partial w_{z}}\right\vert
_{1}&\simeq& \frac{-2(W_{H}-W_{p}-1)}{\sqrt{\pi }}\exp \left[
-(W_{H}-W_{p}-1)^{2}\right] +\nonumber \\
 &+& \frac{2}{\sqrt{\pi }}\frac{n_{1}}{n_{0}}\left( 
\frac{T_{p}}{T_{H}}\right) ^{3/2}\exp \left[ -\frac{T_{p}}{T_{H}}\right]
\label{eqn46}
\end{eqnarray}
Taking now typical values for the above quantities as they prevail
in the LISM interface we can decide whether positive growth
rates are possible. In view of the values we display in Figures \ref{figPParam}
and \ref{figHParam} of Section 1 we select the following typical
values: $W_{H}\simeq 3$; $W_{p}\simeq 0.3$; $T_{p}/T_{H}\simeq
12000/8000=1 $, and thus:
\begin{eqnarray}
\left\vert \frac{\partial f_{p,z}^{\times }}{\partial w_{z}}\right\vert _{1}
&\simeq &\frac{-2(1.7)}{\sqrt{\pi }}\exp [-(1.7)^{2}]+\frac{2}{\sqrt{\pi }}
\frac{n_{1}}{n_{0}}(1.5)^{\frac{3}{2}}{\rm e}^{-\frac{3}{2}}=\   \label{eqn47} \\
&=&\frac{2}{\sqrt{\pi }}\exp [-(1.7)^{2}]\left( \frac{n_{1}}{n_{0}}
(1.5)^{3/2}\exp \left[ +0.2\right] -1\right)  \nonumber
\end{eqnarray}%
meaning that only under unlikely conditions, not met in our calculations
above and characterized by:
\begin{equation}
\frac{n_{1}}{n_{0}}\geq \frac{1}{1.5^{3/2}}\exp(-0.2)=0.44  \label{eqn48}
\end{equation}
a relaxation of the distribution function $f_{p,z}^{\times }$ by excitation
of electron plasma waves is likely to occur.

It thus remains to inspect relaxation processes connected to the
nonlinear interaction of protons with MHD turbulences, amongst which
pitch angle diffusion processes seem to be the most effective ones.
These, however, depend on the local magnetic fields and the existing MHD
turbulence levels. Of course, nothing is known at present about LISM
magnetic fields and turbulences. This is why no good estimate of
pitch-angle diffusion processes is possible here as is otherwise feasible on
the basis of an expression already derived by Hasselmann \& Wibberenz (1970)
\nocite{hasselmann_wibberenz:70} in which the typical time period for pitch
angle diffusion is given by:
\begin{equation}
\tau _{\parallel }\simeq \frac{\lambda _{\parallel }}{v}=\frac{3}{8}
\int_{-1}^{+1}\frac{\left( 1-\mu ^{2}\right) ^{2}d\mu }{D_{\mu \mu }},
\label{eqn49}
\end{equation}
where $\lambda _{\parallel }$ is the mean free path with respect to
pitch-angle scattering, $\mu $ is the pitch angle cosine, and $D_{\mu \mu }$
is the Fokker-Planck pitch angle diffusion coefficient. The latter, however,
is a complicated function of particle velocity, local magnetic field
magnitudes, Alfv{\'e}n velocities, and turbulence levels. This coefficient
$D_{\mu \mu }$ then leads to velocity-dependent mean free paths 
$\lambda _{\parallel }(v)$ and to corresponding scattering periods $\tau
_{\parallel }(v)$ (as given by Chalov \& Fahr 1999a,b
\nocite{chalov_fahr:99a}\nocite{chalov_fahr:99b} for keV-energetic ions). The
needed input information to evaluate the above expression unfortunately are
completely lacking for the LISM interface region and furthermore the newly
injected ions from the H-atom population in this region only have energies
of about 10~eV. Thus we cannot estimate $\tau_{\parallel }$ in this way.

Instead we estimate the relaxation times connected with pitch angle
scattering on the basis of the reasonable expectation that the locally
appearing MHD turbulence levels are not primarily due to convected
pre-existing turbulences, which have not been quantitatively described
in the literature, but due to wave-driving processes of secondary
protons injected via charge exchange processes into an unstable mode of
the proton distribution function. Let us assume that the newly created
protons produced by charge exchange are populated in a region of
velocity space where they represent some free kinetic energy by which
they are able to drive some MHD\ wave power (see, e.g., Huddleston \&
Johnstone 1992\nocite{huddleston_johnstone:92}, Williams \& Zank 1994,
\nocite{williams_zank:94} Fahr \& Chashei 2002\nocite{fahr_chashei:02}).
Let us then assume that this free energy is pumped into turbulent wave
power essentially at an injection wave number $k_{i}$. The following
considerations are of course only relevant if an interstellar magnetic
field is present. For an unmagnetized interstellar medium what follows
here would thus be irrelevant.

Adopting that near the symmetry axis (stagnation line!) a background
magnetic field $B_{LISM}$ exists which is oriented parallel to the
stagnation line (i.e., according to an MHD interaction model with $\vec{B}
_{LISM}\parallel \vec{V}_{LISM}$ as treated, e.g., by Baranov \& Zaitsev
1995\nocite{baranov_zaitsev:95}, Pogorelov \& Matsuda 1998\nocite{pogorelov_matsuda:98},
or Ratkiewicz \& McKenzie 2003\nocite{ratkiewicz_mckenzie:03}), one then
can assume that Alfv\'{e}n waves are
excited by this free energy which then propagate upstream and downstream of the
plasma flow. The free energy of newly injected protons stemming from
the bulk of the H-atom distribution function can then be assumed to be
pumped into the wave field at the injection wave number fulfilling the
resonance conditions and given in the bulk plasma system by:
\begin{equation}
\mp k_{i}\simeq \Omega /\left( v_{A}\pm \left( u_{H}-u_{p}\right) \right)
\label{eqn49a}
\end{equation}

Since hydromagnetic Alfv\'{e}n waves have frequencies $\omega =v_{A}\
k\leq \Omega $ the cyclotron resonance condition can only be met by ions
with positive/negative parallel velocities resonating with waves with
negative/positive wave vectors. In the outer interface plasma newly
injected ions with $u_{H}\geq u_{p}$ have positive parallel velocities
in the plasma frame and thus can only resonate with waves propagating in
the direction opposite to them with wave vector magnitudes $k_{i}\simeq
\Omega /(v_{A}+(u_{H}-u_{p}))$. When resonating with these waves, energy
can be exchanged between such ions and the waves up to an equilibrium
situation. This injected wave power then cascades from $k_{i}$ to larger
wave numbers described by a so-called wave-wave diffusion coefficient
$D_{kk}$ (see Zhou \& Matthaeus 1990\nocite{zhou_matthaeus:90a}) until
the power arrives at a wave number $k_{dis}\simeq \Omega /v_{A}$ where
it is resonantly absorbed by the bulk of the protons.

We now determine the time $\tau _{kk}$ needed to cascade the
injected power to the dissipation wave scale -- in order thereby
to derive the time needed to transfer energy from the secondary
proton hump in the proton distribution function to the primary proton bulk.
Following Chashei et al. (2003)\nocite{chashei_etal:03} we obtain:
\begin{equation}
\tau _{kk}\simeq \frac{\left( \Delta k\right) ^{2}}{4D_{kk}}=\frac{\left(
k_{i}-k_{dis}\right) ^{2}}{4D_{kk}}=\frac{\left( \frac{\Omega }{
v_{A}+(u_{H}-u_{p})}-\frac{\Omega }{v_{A}}\right) ^{2}}{4D_{kk}}
\label{eqn50}
\end{equation}
According to Zhou \& Matthaeus (1990)\nocite{zhou_matthaeus:90a}, we can
represent the k-space diffusion coefficient by the following expression:
\begin{equation}
D_{kk}=Cv_{A}k_{i}^{7/2}\sqrt{\frac{4\pi W_{i}}{B_{LISM}}}  \label{eqn51}
\end{equation}
where $W_{i}$ denotes the spectral power at wave number $k_{i}$, and $C
\simeq 0.1$ is a constant. Assuming that wave energy is injected at the same
rate as it cascades down to $k_{dis}$, we can determine $W_{i}$ by
the following relation:
\begin{equation}
W_{i}\Delta k\simeq \beta _{ex}\Delta \epsilon _{i}\tau _{kk}=\beta
_{ex}\Delta \epsilon _{i}\frac{\Delta k^{2}}{4D_{kk}}  \label{eqn52}
\end{equation}
where $\beta _{ex}=n_{p}n_{H}\sigma _{ex}V_{rel}$ is the mean charge
exchange rate between H-atoms and protons, and $\Delta \epsilon
_{i}=\epsilon _{f}\left( \frac{1}{2}m_{p}\left( u_{H}-u_{p}\right)
^{2}\right) $ is the free energy pumped from the newly injected protons into
the wave field. With Eqs. (\ref{eqn50}) through (\ref{eqn52}) one thus finds:
\begin{equation}
W_{i}^{3/2}\simeq \frac{B_{LISM}}{2Cv_{A}k_{i}^{7/2}\sqrt{\pi }}\beta
_{ex}\Delta \varepsilon _{i}\Delta k=\frac{\sqrt{\rho _{p}}\beta _{ex}\Delta
\epsilon _{i}\Delta k}{Ck_{i}^{7/2}}  \label{eqn53}
\end{equation}
With the above result we finally find the diffusion time $\tau _{kk}$ as
given by:
\begin{eqnarray}
\tau _{kk}&\simeq& \frac{W_{i}\Delta k}{\beta _{ex}\Delta \epsilon _{i}}=\frac{
\Delta k}{\beta _{ex}\Delta \epsilon _{i}}\left( \frac{\sqrt{\rho _{p}}\beta
_{ex}\Delta \epsilon _{i}\Delta k}{4Ck_{i}^{7/2}}\right) ^{2/3}= \\
 &=&\frac{\left( 
\frac{u_{H}-u_{p}}{v_{A}}\right) ^{5/3}}{2C^{2/3}\left( n_{H}~\sigma
_{ex}V_{rel}~\epsilon _{f}\left( u_{H}-u_{p}\right) ^{2}k_{i}^{2}\right)
^{1/3}} \nonumber \label{eqn54}
\end{eqnarray}

Let us now assume for estimation purposes that $V_{rel}$ can be reasonably
well approximated by $V_{rel}\simeq (u_{H}-u_{p})$. Then we arrive at the
following expression:
\begin{eqnarray}
\tau _{kk}&\simeq& \frac{\left( u_{H}-u_{p}\right) ^{2/3}}{2C^{2/3}v_{A}^{5/3}
\left( n_{H}~\sigma _{ex}~\epsilon _{f}~k_{i}^{2}\right) ^{1/3}}= \\
&=&\frac{
\left( u_{H}-u_{p}\right)^{2/3}\left( v_{A}+u_{H}-u_{p}\right) ^{2/3}}{
2\epsilon _{f}^{1/3}C^{2/3~}v_{A}^{5/3}~\Omega ^{2/3}\left( n_{H}~\sigma
_{ex}\right) ^{1/3}} \nonumber \label{eqn55}
\end{eqnarray}
For further evaluation of this expression we transform it into the following
form:
\begin{equation}
\tau _{kk}\simeq \frac{\left( X_{H}-X_{p}\right) ^{2/3}\left(
1+X_{H}-X_{p}\right) ^{2/3}}{2\epsilon _{f}^{1/3}C^{2/3}\Omega ^{2/3}\left(
v_{A}~n_{H}~\sigma _{ex}\right) ^{1/3}}  \label{eqn56}
\end{equation}
where the quantities $X_{H}$ and $X_{p}$ denote the corresponding bulk
velocities normalized by $v_{A}$.

It remains to find some adequate value for the fraction $\epsilon _{f}$ of
energy that is transferred as free energy to the wave fields. This value we
intend to derive in analogy to a similar derivation given by
Chashei et al. (2003)\nocite{chashei_etal:03} for the free energy of
newly injected pick-up ions in the inner heliosphere. The following
estimation cannot be carried out prior to at least some qualitative view of
the configuration and magnitude of the magnetic field $B_{LISM}$ in the
outer interface.

Hence we assume here that close to the stagnation line on the
interstellar side of the interface we have a field $\vec{B}_{{\rm 
LISM}}$ which is parallel to the plasma flow, i.e., parallel to $\vec{u}_{p}$
(see, e.g., the model presented by Baranov \& Zaitsev 1995\nocite
{baranov_zaitsev:95}). Then newly ionized H-atoms are implanted in the
proton distribution function roughly at a velocity $\Delta u=\left\vert 
\vec{u}_{H}-\vec{u}_{p}\right\vert $ relative to the proton bulk flow. Injection
of a newly created proton from the H-atom bulk on the average will eventually
lead to the following relative velocity $g$ with respect to the proton bulk:
\begin{equation}
g\simeq \sqrt{\Delta u^{2}+c_{H}^{2}}  \label{eqn57}
\end{equation}
where $c_{H}$ is the thermal velocity of the H-atoms.

Now we consider the two frames of Alfv\'{e}n waves moving with the velocity 
$\pm v_{A}$ relative to the proton bulk frame in up-field and down-field
directions. With respect to these wave frames the injected protons have
relative velocities given by:
\begin{equation}
g_{A\pm }=\sqrt{\left( v_{A}\pm \Delta u\right) ^{2}+c_{H}^{2}}=\sqrt{
v_{A}^{2}\pm 2v_{A}~\Delta u+g^{2}}  \label{eqn58}
\end{equation}
and by resonance with the up-field waves will rapidly be pitch-angle
scattered onto the accessible spherical shell that is associated with the
upgoing wave-frame (see Huddleston \& Johnstone 1992\nocite{huddleston_johnstone:92}). With respect to the proton bulk frame they
thereby lose an energy $\Delta \epsilon
_{f}$, taken as the pitch-angle average, given by (see Fahr \& Chashei 2002\nocite{fahr_chashei:02}, Chashei et al.
2003\nocite{chashei_etal:03}):
\begin{equation}
\Delta \epsilon _{f}\simeq \frac{1}{2}m_{p}\left[ g^{2}-\left\langle
g_{+}^{2}\left( \vartheta \right) \right\rangle \right] =\epsilon _{f}\left( 
\frac{1}{2}m_{p}~g^{2}\right)  \label{eqn59}
\end{equation}
where the angle-averaged velocity $\left\langle g_{+}^{2}(\vartheta
)\right\rangle $ of the pitch-angle scattered new protons is given by:
\begin{equation}
\left\langle g_{+}^{2}\left( \vartheta \right) \right\rangle =
\frac{\int\limits_{\cos \vartheta _{\max }^{\pm}}^1
\left( v_{A}^{2}+g_{A+}^{2}-2v_{A}~g_{A\pm }\cos \vartheta \right)
d\cos \vartheta}{ 1-\cos\vartheta _{\max }^{\pm }} \label{eqn60}
\end{equation}
Here the maximum permitted inclination angles are defined by:
\begin{equation}
\cos \vartheta _{\max }^{+}=\frac{v_{A}+\Delta u}{g_{A+}}  \label{eqn61}
\end{equation}
\ After evaluation of the integral in the above expressions one then finds:
\begin{equation}
\left\langle g_{+}^{2}\left( \vartheta \right) \right\rangle
=2v_{A}^{2}-2v_{A}~\Delta u+g^{2}-\frac{v_{A}~c_{H}^{2}}{\sqrt{
v_{A}^{2}+g^{2}-2v_{A}\Delta u}}  \label{eqn62}
\end{equation}
which leads to:
\begin{eqnarray}
\Delta \epsilon _{f+}&\simeq& \frac{1}{2}m_{p}\left[ g^{2}-\left\langle
g_{+}^{2}\left( \vartheta \right) \right\rangle \right] =\\
 &=&-\frac{1}{2}
m_{p}g^{2}\left( 2z_{A}^{2}-2z_{A}z_{U}-\frac{z_{A}z_{H}^{2}}{\sqrt{
1+z_{A}^{2}-2z_{A}z_{U}}}\right) \nonumber \label{eqn63}
\end{eqnarray}
where the following notations have been used: $z_{A}=v_{A}/g$; 
$z_{H}=c_{H}/g$; and $z_{U}=\Delta u/g$. If we now put into the
expressions above some concrete numbers, we find:
\begin{equation}
v_{A}=v_{A0}\left( B_{LISM}/B_{0}\right) \sqrt{n_{0}/n_{LISM}}  \label{eqn64}
\end{equation}
where the quantities with index ``0'' are
reference quantities and, taking these reference values for the solar wind
at the orbit of the earth, i.e., $v_{A0}=40$~km/s, $B_{0}=5$~Gamma and 
$n_{0}=5$~cm$^{-3}$, we finally obtain:
\begin{equation}
v_{A}=v_{A0}(3/50)\sqrt{5/0.1}\simeq 17~{\rm km/s}  \label{eqn66}
\end{equation}
Then taking $U_{H}\simeq 20$~km/s, $U_{p}\simeq 10$~km/s, and $c_{H}\simeq
15 $~km/s, we obtain:

$g\simeq \sqrt{\Delta U^{2}+c_{H}^{2}}=\sqrt{10^{2}+15^{2}}$~km/s $=\sqrt{
325}$~km/s $=18$~km/s;
$z_{A}=v_{A}/g=17/18=0.94$, $z_{U}=\Delta U/g=0.55$; $z_{H}=c_{H}/g=0.83$;
$X_{H}=U_{H}/v_{A}=20/17=1.18$ ; $X_{p}=U_{p}/v_{A}=10/17=0.58$
and thus we find:
\begin{eqnarray}
\Delta \epsilon _{f\pm }&=&-\frac{1}{2}m_{p}g^{2}\left( 2\cdot 0.94^{2}-2\cdot
0.94\cdot 0.55-\right.\\
&-&\left. \frac{0.94\cdot 0.83^{2}}{\sqrt{1+0.94^{2}-2\cdot 0.94\cdot
0.55}}\right) =0.55  \nonumber \label{eqn67}
\end{eqnarray}
which due to the fact that newly created protons only resonate with waves
propagating in the up-field direction finally yields $\epsilon _{f+}\simeq 0.55$.
Hence one obtains:
\begin{equation}
\tau _{kk}\simeq \frac{\left( 1.18-0.58\right) ^{2/3}\left(
1+1.18-0.58\right) ^{2/3}}{2\cdot \epsilon _{f+}^{1/3}C^{2/3}\Omega
^{2/3}\left( v_{A}~n_{H}~\sigma _{ex}\right) ^{1/3}}  \label{eqn68}
\end{equation}
leading to:
\begin{equation}
\tau _{kk}\simeq \frac{\left( 0.6\right) ^{2/3}\left( 1.6\right) ^{2/3}}{
2\cdot 0.55^{1/3}C^{2/3}\Omega ^{2/3}\left( n_{H~}v_{A}\sigma _{ex}\right)
^{1/3}}  \label{eqn69}
\end{equation}
Putting in numbers like: $C=0.1$; $n_{H}=0.1$; $\sigma _{ex}=2.5\cdot
10^{-15}$~cm$^{2}$; $\Omega =9.5\cdot 10^{-2}(B_{LISM}/B_{0})=5.7\cdot
10^{-3}$~s we then obtain:
\begin{eqnarray}
\tau _{kk}&=& \frac{0.71 \cdot 1.36}{2\cdot 0.82\cdot 0.22\cdot 0.033\left(
1.6\cdot 10^{6}\cdot 2.5\cdot 10^{-16}\right) ^{1/3}}\simeq \nonumber \\
&\simeq& 10^{4}~{\rm s}
\label{eqn70}
\end{eqnarray}
This now means that the isotropization and redistribution of newly
incorporated protons from the primary ring distribution to the
associated spherical shell distribution will occur within typical time
periods of a few hours. This, however, does not mean that a full
relaxation towards an associated equilibrium distribution is likely to
occur within a similar time period. The resulting distribution function
$f$ is still far from fulfilling the requirement $\delta \left( \int
f~\ln f~d^{3}v\right) /\delta t=0$. One other point, however, becomes
very clear when comparing the period $\tau_{kk}$ with the period
$\tau_{p}$ yielding $\tau_{p}/\tau_{kk}\approx 10^{4}$, namely: 
the process of wave driving operates much faster than the proton-proton 
relaxation, so that at least the fraction $\varepsilon _{f}$ of the
kinetic energy of newly implanted ions is pumped into wave turbulence.

\section{Conclusions}

In purely hydrodynamic multi-fluid interaction codes describing the
interaction of the partially ionized interstellar medium with the solar
wind plasma, both the proton fluid and the H-atom fluid are generally
represented by the moments of distribution functions which are taken to
be bulk-flow shifted Maxwellians. We have shown for the region close to
the stagnation line, outside of the heliopause but inside of the outer
bowshock, that this assumption is substantially violated for the H-atom
flow, while it is only mildly violated for the proton flow. This result,
however, is due to the values of interstellar parameters adopted in our
study, where we have taken the LISM H-atom density = LISM proton density
= 0.1~cm$^{-3}$. In that case the p-p Coulomb relaxation processes are
quite effective in keeping the resulting proton distribution function
close to a shifted Maxwellian.

For different LISM parameter combinations, e.g., for lower LISM proton
densities, or at different places of the heliospheric interface, the
assumption of having the proton distribution function close to a shifted
Maxwellian may be violated. This is because the Coulomb relaxation
period $\tau_p \simeq T_p^{3/2}/n_p$ which means that for lower proton
densities and higher proton temperatures this relaxation period may
easily increase to values $\tau_p >> \tau_{ex}$ and thus induce
substantial deviations of the resulting proton distribution
function from a shifted Maxwellian. For instance, inside the heliopause,
where very low proton densities and high proton temperatures
prevail, substantial deviations of the proton distribution from
shifted Maxwellians should be expected unless alternative relaxation
processes operate, as discussed in Section 4.

\begin{acknowledgements}
We thank Dr. K. Scherer for supplying us with computer outputs run
on the basis of the multi-fluid interaction model published by
Fahr et al. (2000).\nocite{fahr_etal:00}
M. Bzowski gratefully acknowledges hospitality of the Institute of
Astrophysics and Extraterrestrial Research, Bonn, Germany (AIUB), where a
major part of the work was carried out. This research was
supported by the DFG/PAS Cooperative Project 436 POL 113/80/0 and by the Polish
State Committee for Scientific Research grant No 2P03C 005 19.
\end{acknowledgements}

\bibliographystyle{aa}
\bibliography{0599}
\end{document}